\documentclass[aps,twocolumn,superscriptaddress,preprintnumbers]{revtex4}
\usepackage{braket}
\usepackage{subfigure}
\usepackage[latin9]{inputenc}
\usepackage{color}
\usepackage{verbatim}
\usepackage{amsthm,amsmath,amssymb}
\usepackage{mathrsfs}
\usepackage{graphicx}
\usepackage{esint}
\usepackage{mathrsfs}
\usepackage{bm}

\usepackage{hyperref}

\makeatletter
\@ifundefined{textcolor}{}
{%
 \definecolor{BLACK}{gray}{0}
 \definecolor{WHITE}{gray}{1}
 \definecolor{RED}{rgb}{1,0,0}
 \definecolor{GREEN}{rgb}{0,1,0}
 \definecolor{BLUE}{rgb}{0,0,1}
 \definecolor{CYAN}{cmyk}{1,0,0,0}
 \definecolor{MAGENTA}{cmyk}{0,1,0,0}
 \definecolor{YELLOW}{cmyk}{0,0,1,0}
}

\usepackage{mathrsfs}

\draft
\makeatother

\begin{document}

\title{Parafermions with symmetry-protected non-Abelian statistics}

\author{Jian-Song Hong}
\affiliation{International Center for Quantum Materials and School of Physics, Peking University,
Beijing 100871, China}
\affiliation{Hefei National Laboratory, Hefei 230088, China}
\author{Su-Qi Zhang}
\affiliation{School of Physics and Institute for Quantum Science and Engineering, Huazhong University
of Science and Technology, Wuhan 430074, China}
\affiliation{Hefei National Laboratory, Hefei 230088, China}
\author{Xin Liu}
\affiliation{School of Physics and Institute for Quantum Science and Engineering, Huazhong University
of Science and Technology, Wuhan 430074, China}
\affiliation{Hefei National Laboratory, Hefei 230088, China}
\author{Xiong-Jun Liu}
\thanks{Corresponding author: xiongjunliu@pku.edu.cn}
\affiliation{International Center for Quantum Materials and School of Physics, Peking University,
Beijing 100871, China}
\affiliation{Hefei National Laboratory, Hefei 230088, China}
\affiliation{International Quantum Academy, Shenzhen 518048, China}

\begin{abstract}
Non-Abelian anyons have garnered extensive attention for obeying exotic non-Abelian statistics and having potential applications to fault-tolerant quantum computing. While the prior research has predominantly focused on non-Abelian statistics without the necessity of symmetry protection, recent progresses have shown that symmetries can play essential roles and bring a notion of the symmetry-protected non-Abelian (SPNA) statistics.
In this work, we extend the concept of SPNA statistics to strongly-correlated systems which host parafermion zero modes (PZMs). This study involves a few fundamental results proved here. First, we unveil a generic unitary symmetry mechanism that protects PZMs from local couplings. Then, with this symmetry protection, the PZMs can be categorized into two nontrivial sectors, each maintaining its own parity conservation, even though the whole system cannot be dismantled into separate subsystems due to nonlinear interactions. Finally, by leveraging the parity conservation of each sector and the general properties of the effective braiding Hamiltonian, we prove rigorously that the PZMs intrinsically obey SPNA statistics. To further confirm the results, we derive the braiding matrix at a tri-junction. We also propose a correlated quantum nanowire model that accommodates a pair of PZMs protected by mirror symmetry and satisfying the generic theory.
This work shows a broad spectrum of strongly-correlated systems capable of hosting fractional SPNA quasiparticles and enriches our comprehension of fundamental quantum statistics linked to the symmetries that govern the exchange dynamics.

\end{abstract}

\maketitle


\section{Introduction}
Quantum statistics of indistinguishable particles sets a fundamental principle in quantum mechanics~\cite{Shankarbook}, which classifies the elementary and quasi-particles into bosons, fermions, and anyons according to different basic rules upon their exchange~\cite{Anyon1977,Wilczek1982,WilczekBook}. 
The non-Abelian statistics adds to a captivating notion in the anyon statistics, which revolutionizes the traditional dichotomy of particles into bosons and fermions, triggering the paradigm shifts in both fundamental physics and potential applications~\cite{Nayak1996,Ivanov2001,DasSarma2005,Nayak2008,Alicea2011}. Rather than a global statistical phase featuring the Abelian quantum statistics, the exchange of non-Abelian anyons is characterized by matrices which transform the quantum states of the many-body system. Such braiding statistics can be construed as a non-Abelian geometric phase defined in the degenerate ground states~\cite{Berry1997,GeometricNRP2019}, and has driven the broad exploration of the exotic non-Abelian topological orders~\cite{SternReview2010}. The quantum information can be encoded non-locally in non-Abelian anyons and manipulated by topological braiding with robustness to local noises, opening an elegant avenue for fault-tolerant quantum computation~\cite{Kitaev2003,PachosBook,Pachos2017}.

Among various non-Abelian anyons, the Majorana zero modes (MZMs) stand out due to the recent considerable experimental progress in the topological superconductors~\cite{Mourik2012,MTDeng2012,Rokhinson2012,Shtrikman2012,JFJia2012, Marcus2013,JFJia2014,Yazdani2014,Marcus2015,Marcus2016,Molenkamp2016, Molenkamp2017,HDing2018a,HDing2018b,Marcus2019, Molenkamp2019,Yazdani2019,Zhang2022,Microsoft2023}, where the Majorana modes emerge as quasiparticles identical to their antiparticles~\cite{Kitaev2001,AliceaRPP,Flensberg2012,Beenakker2013, Franz2015,Sato2017,YPHe2020}. Two MZMs define a non-local complex fermion. The qubits encoded by these non-local fermions are robust against local perturbations~\cite{Kitaev2001}. Braiding and measuring MZMs can implement Clifford gates, although these are not sufficient for universal quantum computation~\cite{Gottesman1998}. From an algebraic standpoint, MZMs are considered as a special case within the broader category of $\mathbb{Z}_{N}$ parafermion zero modes (PZMs) with $N=2$~\cite{FradkinKadanoff,Fendley2012,Motruk2013topological,Bondesan2013,
Meidan2017,GMZhang2018a}. 
Physically, the parafermions are fractionalized MZMs, emerging only in the strongly correlated systems~\cite{SternPRX2012,MCheng2012,AliceaNC2013,Barkeshli2013,
Barkeshli2014,Stern2014,Loss2014a,Loss2014b,Loss2014c,Oreg2017general,
Oreg2017chiral,Loss2019,Santos2020,Loss2020bilayer,Oreg2020predicted,
Silva2022,Gefen2022}. The braiding phases of PZMs also exhibit fractionalization compared to those of MZMs, manifesting a kind of non-Abelian fractional statistics. The parafermions can encode the nonlocal topological qudits~\cite{FockPara}, and braiding parafermion brings us closer to the computational universality than braiding MZMs~\cite{AliceaNC2013,Loss2016}.

The non-Abelian statistics was initially introduced to singly existing MZMs which necessitate no symmetry protection~\cite{Nayak1996,Ivanov2001,Alicea2011}. The presence of symmetries can enrich the broad classes of topological superconductors~\cite{SCZhang2009,TeoKane2010,Timm2010,Beenakker2011, KTLaw2012,Nagaosa2012,KaneMele2013,Berg2013,DLoss-a,DLoss-b,Nagaosa2014,magneticTSC,Oreg2019,mirrorTSC,Ueno2013,Sato2014crystalline,
C4TSC,InversionTSC}, in which a single topological defect may bind multiple zero modes and generate the ground state degeneracy. 
It was first shown in the time-reversal invariant topological superconductors that the Majorana Kramers pairs obey non-Abelian statistics due to the protection of time-reversal symmetry, rendering a new family of quantum statistics called the symmetry-protected non-Abelian (SPNA) statistics~\cite{XJL2014PRX,Gao2016}. This prediction shows an insight into the nontrivial roles of symmetries in the quantum statistics. The SPNA statistics has been further considered in various symmetry-protected topological superconductors~\cite{Sato2014mirror,JSH2022,YWuReview2023,
NittaReview2024}, and is also the essential mechanism~\cite{JSH2022} underlying the braiding statistics of Dirac zero modes~\cite{Nitta2012dirac,Loss2013fractional,YWu2020a,YWu2020b,YWu2022}. 
The unitary and anti-unitary symmetries are fundamentally different in protecting the non-Abelian statistics. For the system with anti-unitary symmetry, e.g. the time-reversal invariant topological superconductor, the non-Abelian statistics of Majorana Kramers pairs requires that not only the static topological superconductor satisfy the time-reversal symmetry, but also the braiding operation, which characterizes a dynamical evolution, satisfy a time-reversal symmetry individually~\cite{Gao2016}. This is because the time-reversal symmetry for the static superconductor cannot exclude the local coupling between two MZMs of a single Kramers pair in the generic dynamical evolution~\cite{localmixing1,localmixing2,Knapp2020} due to the profound dynamical symmetry-breaking of anti-unitary symmetries~\cite{Gao2016,JasonAlicea2020}. The dynamical symmetry breaking with coupling to environment opens up an interesting issue on the robustness of time-reversal invariant topological phases in the open systems~\cite{Cooper2020,Zhai2021,Cai2021}. In contrast, the Majorana modes protected by unitary symmetries are intrinsic SPNA anyons without suffering the dynamical symmetry breaking. In particular, when the unitary symmetry is non-Abelian, the entire system cannot be block diagonalized into decoupled sectors corresponding to different symmetry eigenvalues, but the non-Abelian statistics is fully protected by the symmetry~\cite{JSH2022}. 
The SPNA statistics elucidates that the quantum statistics of indistinguishable (quasi)-particles is essentially a dynamical rather than a static feature. In additional to the fundamental relevance, the symmetry-protected MZMs enrich new schemes for topological quantum computing~\cite{JSH2022,OregReview2019,MCheng2021,TanakaMKP2022,LFu2022}.

Topological superconductors belong to the free-fermion topological phases~\cite{XGWen2012}, for which the SPNA statistics has been so far proposed and thoroughly studied. Whether or not this concept can be generalized to strongly correlated systems, in particular those with PZMs, is an open question with fundamental importance, but confronts formidable challenges. 
First, for parafermions it was not even known how to generally introduce the symmetry protection. Further, due to strong interactions, systems accommodating PZMs cannot be dismantled into separate symmetry sectors, bringing the challenges for generic study. 
Finally, the braiding transformation of PZMs is nonlinear~\cite{Loss2016}, and the characterization with symmetries can be even more complicated. 
Nevertheless, generalizing the SPNA statistics to strongly correlated systems could bring a new horizon in the fundamental physics of quantum statistics and promote intriguing applications to quantum computing. 

\begin{figure}
\centering
\includegraphics[width=\columnwidth]{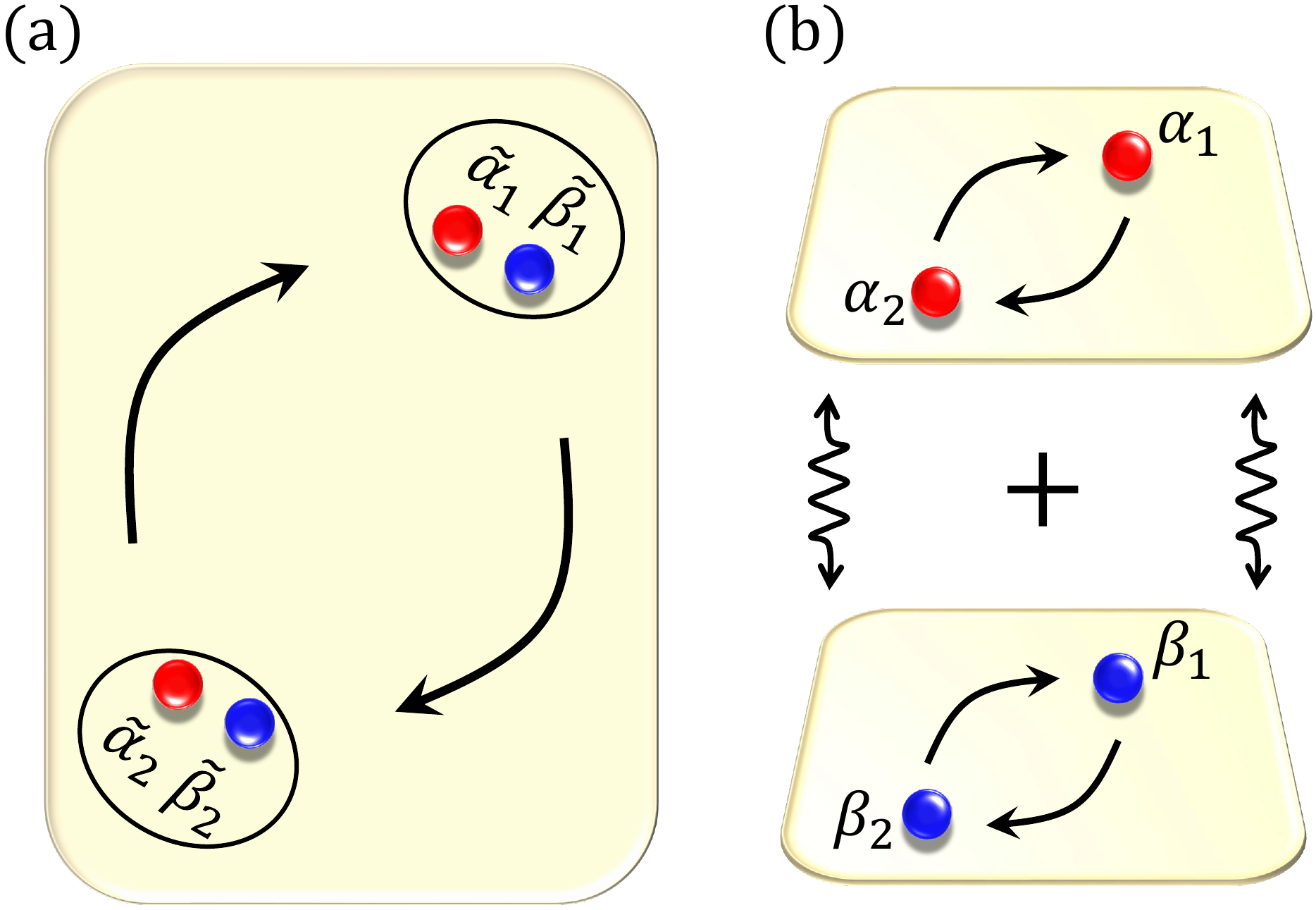}
\caption{Schematic diagrams for two pairs of PZMs residing at two sides of the system protected by a unitary symmetry. (a) The braiding of two pairs of PZMs reduces to (b) two individual sectors (the PZMs are redefined according to a sophisticated unitary transformation $\tilde\alpha_i,\tilde\beta_i\rightarrow\alpha_i,\beta_i$), each of which braids two PZMs independently. The wavy lines denote the interaction between two subsystems. These terms indicate that although the PZMs are categorized into two independent sectors, the subsystems are not independent ones.}
\label{MainResult}
\end{figure}

In this work, we propose the theory of non-Abelian statistics for parafermions protected by unitary symmetry, and present a systematic study consisting of the profound results. We first unveil the generic symmetry mechanism for protecting the PZMs, as formulated with a unitary transformation on the anyon parities, with which the local couplings are prohibited. With this symmetry protection, we show that the PZMs can be categorized into two nontrivial symmetric sectors, each maintaining its own parity conservation, even though the whole system cannot be dismantled into separate subsystems due to nonlinear interactions. By harnessing the parity conservation of each sector and leveraging the fundamental properties of the effective braiding Hamiltonian, we prove rigorously that the PZMs intrinsically obey SPNA statistics, as captured by two independent braiding operations [see Fig.~\ref{MainResult}]. 
We further propose a physical model that accommodates a pair of PZMs protected by a mirror symmetry and satisfying the theory of SPNA statistics.
The strong correlation effects are crucial for the model realization and affect the concrete form of the braiding matrix, but the SPNA statistics are generally obtained. 

The rest of the article is organized as follows. In Sec.~\ref{SecProof}, we provide a rigorous proof of the SPNA statistics of PZMs based on showing several theorems. Then in Sec.~\ref{SecJunction}, we derive the braiding matrix through a tri-junction, which further confirms the SPNA statistics of the parafermions. A concrete physical realization is proposed in Sec.~\ref{SecModel} through a strongly correlated nanowire model hosting PZMs with mirror symmetry. The conclusion and discussions are given in Sec.~\ref{SecEnd}, with the important future issues being commented.

\section{Non-Abelian braiding of PZMs with unitary symmetry protection}\label{SecProof}

In this section, through a systematic study we prove rigorously that PZMs with unitary symmetry protection are intrinsic non-Abelian fractional quasiparticles. We first unveil the generic mechanism of unitary symmetry protection for the locally coexisting PZMs, for which all the local couplings are forbidden. This result leads to two key consequences. One is that the PZMs can be categorized into two separate sectors by redefining these modes from a sophisticated symmetric decomposition, yet the whole system cannot be decoupled into independent subsystems due to the strong interactions. The other is that in terms of the decomposed PZMs, the anyon parity of each symmetry sector is conserved. With the parity conservation and topological spin statistics theorem, we prove rigorously that the PZMs with unitary symmetry protection intrinsically obey the SPNA statistics.

\subsection{Unitary symmetry protection}

We first introduce the system of locally-coexisting PZMs and determine the generic form of the local coupling terms breaking symmetry protection. Consider two pairs of $\mathbb{Z}_{N}$ PZMs localized at two edges (domain walls) of the system with the commutation relations
\begin{eqnarray}
\tilde{\alpha}_{i}\tilde{\beta}_{j} & = & \omega\tilde{\beta}_{j}\tilde{\alpha}_{i},\\
\tilde{X}_{i}\tilde{X}_{j} & = & \omega^{\text{sgn}(j-i)}\tilde{X}_{j}\tilde{X}_{i},\ \tilde{X}=\tilde{\alpha},\tilde{\beta},
\end{eqnarray}
where $\omega=\exp(2\pi i / N)$, $\tilde{\alpha}_{i=1,2}$, $\tilde{\beta}_{j=1,2}$ are unitary operators that commute with the system's Hamiltonian and satisfy $\tilde{\alpha}_{i}^{N}=\tilde{\beta}_{i}^{N}=1$, and $\tilde{\alpha}_{1(2)}$, $\tilde{\beta}_{1(2)}$ localize at the left (right) edge of the system. Without symmetry protection, the PZMs $\tilde{\alpha}$ and $\tilde{\beta}$ of different flavours may be coupled together by local Hamiltonian $H_{\text{c}}=\sum_{k,l}\xi_{k,l}\tilde{\alpha}_{1}^{k}\tilde{\beta}_{1}^{l}+\sum_{m,n}\eta_{m,n}\tilde{\alpha}_{2}^{m}\tilde{\beta}_{2}^{n}$, where $\xi_{k,l}^{*}=\omega^{kl}\xi_{N-k,N-l}$ and $\eta_{m,n}^{*}=\omega^{mn}\eta_{N-m,N-n}$ for hermitianity. Since the coupling terms must conserve the total parity defined by $\tilde{Q}_{\text{tot}}=\omega\tilde{\alpha}_{1}\tilde{\alpha}_{2}^{\dagger}\tilde{\beta}_{1}\tilde{\beta}_{2}^{\dagger}$, we can write down the concrete form of the local couplings as
\begin{equation}
H_{\text{c}}=\sum_{m=1}^{N-1}\xi_{m}\tilde{Q}_{1}^{m}+\sum_{n=1}^{N-1}\eta_{n}\tilde{Q}_{2}^{n},\label{LocalH}
\end{equation}
where $\xi_{m}=\xi_{N-m}^{*}$ and $\eta_{n}=\eta_{N-n}^{*}$ for hermitianity and $\tilde{Q}_{i}=\omega^{(N+1)/2}\tilde{\alpha}_{i}\tilde{\beta}_{i}^{\dagger}$ denotes the local $\mathbb{Z}_{N}$ parity. The Hamiltonian Eq.~(\ref{LocalH}) describes the generic local coupling terms in the absence of symmetry protection other than the intrinsic total parity conservation. The presence of the these terms pushes the parafermion modes out of zero energy, and the role of symmetry protection is then to forbid these terms as detailed below.

We show that the generic unitary symmetry protection for any pair of PZMs can be expressed through the unitary transformation on the local anyon parities. If there is a unitary symmetry $S$ protecting the PZMs, it implies that the local coupling of these modes are prohibited by requiring that $[H_{\text{c}},S]\neq0$ for arbitrary $H_{\text{c}}$. 
When $N=2$, $\tilde{\alpha}_{i}$ and $\tilde{\beta}_{i}$ are reduced to MZMs. Then $H_c$ only involves the terms $i\xi_{1}\tilde{\alpha}_{1}\tilde{\beta}_{1}$ and $i\eta_{1}\tilde{\alpha}_{2}\tilde{\beta}_{2}$, which can be eliminated by introducing a unitary symmetry $S i\tilde{\alpha}_{i}\tilde{\beta}_{i} S^{-1}=-i\tilde{\alpha}_{i}\tilde{\beta}_{i}$. For PZMs with $N>2$, the symmetry protection mechanism is generally more complicate. To facilitate the derivation, we work in the bases of eigenstates $|q\rangle$ of local parity operator $\tilde{Q}_{i}$. The generic condition $[H_{\text{c}},S]\neq0$ implies that the unitary symmetry $S$ must be a cyclic permutation of the eigenstates with $N$ elements provided that $S|q\rangle$ is still a local parity eigenstate. In this case, there exists a unitary matrix $G$ such that $GSG^{-1}=e^{i\vartheta/N}V^{p}$, where $p$ and $N$ are coprime, $\vartheta$ is a phase factor and $V$ is the shift matrix $V|q\rangle=|q-1\rangle$ (the details are provided in Appendix~\ref{appenA}). This observation can finally be rewritten in the form
\begin{equation}
SQ_{i}S^{-1}=\omega^{p}Q_{i},\label{UnitaryS}
\end{equation}
where $Q_i=G^{-1} \tilde{Q}_i G$ is the local parity operator of PZMs $\alpha'_i=G^{-1} \tilde{\alpha}_i G$ and $\beta'_i=G^{-1} \tilde{\beta}_i G$. The Eq.~(\ref{UnitaryS}) suggests that the symmetry can advance the new local parity $Q_i$ by $p$-increments, where $p$ and $N$ are coprime. An intuitive case is for $N=2$, in which the symmetry operation switches fermion parity between even and odd, which is a familiar feature for MZMs.

\subsection{Parity conservation}

In this subsection, we show a pivotal result that with the generic unitary symmetry protection, as given in Eq.~(\ref{UnitaryS}), the PZMs can be classified into two nontrivial symmetry sectors after a sophisticated transformation on the parafermion modes, with the parity of each sector being conserved. From Eq.~(\ref{UnitaryS}) we can show that the unitary symmetry $S$ transforms the PZMs in the way (see Appendix~\ref{appenB} for more details)
\begin{eqnarray}
S\alpha'_{i}S^{-1} & = & \omega^{p_{1}}\alpha'_{i}e^{i\sum_{n=1}^{N-1}\mu_{n}Q_{i}^{n}},\\
S\beta'_{i}S^{-1} & = & \omega^{p_{2}}\beta'_{i}e^{i\sum_{n=1}^{N-1}\mu_{n}Q_{i}^{n}},
\end{eqnarray}
where $\mu_{n}^{*}=\mu_{N-n}$, $p_1$, $p_2$ are integers that satisfy $p_{1}-p_{2}=p$, and $\alpha'_i=G^{-1} \tilde{\alpha}_i G$, $\beta'_i=G^{-1} \tilde{\beta}_i G$. We see that the symmetry operator mixes $\alpha'_i$ and $\beta'_i$ in general. We can simply the formula by defining a new set of PZMs via the unitary transformation
\begin{eqnarray}
\alpha_{i}=W\alpha'_{i}W^{-1}, \ \beta_{i}=W\beta'_{i}W^{-1},
\end{eqnarray}
where $W=\exp\bigr({i\sum_{n=1}^{N-1}\nu_{n}Q_{i}^{n}}\bigr)$, $\nu_{n}=-\mu_{n}/[(\omega^{n}-1)(\omega^{pn}-1)]$. The generic symmetry protection can be finally rewritten in a decoupled form
\begin{eqnarray}
S\alpha_{i}S^{-1}  =  \omega^{p_{1}}\alpha_{i},\quad
S\beta_{i}S^{-1}  =  \omega^{p_{2}}\beta_{i}.\label{decoupled}
\end{eqnarray}
This is an important result that the PZMs can always be written in terms of eigenmodes of the symmetry, even though the parent topological phase hosting such modes is strongly correlatd and cannot be block diagonalized.  
For the noninteracting phase with MZMs at $N=2$, the above procedure can be easily understood. Namely, for a pair of MZMs $\alpha'_{i}$ and $\beta'_{i}$, the symmetry acts as $S\alpha'_{i}S^{-1} = \pm(\cos\phi\alpha'_{i}+\sin\phi\beta'_{i})$, $S\beta'_{i}S^{-1} = \mp(-\sin\phi\alpha'_{i}+\cos\phi\beta'_{i})$, which give $\mu_{1}=\phi$. Setting $\nu_{1}=-\phi/4$, we have $\alpha_{i}=\cos\frac{\phi}{2}\alpha'_{i}+\sin\frac{\phi}{2}\beta'_{i}$, $\beta_{i}=-\sin\frac{\phi}{2}\alpha'_{i}+\cos\frac{\phi}{2}\beta'_{i}$, and it's easy to verify that $S\alpha_{i}S^{-1} = \pm\alpha_{i}$, $S\beta_{i}S^{-1} = \mp\beta_{i}$. 
Nevertheless, for the parafermion modes, the generic unitary symmetry transformation $W$ given above is highly nonlinear.

The decoupled form of unitary symmetry protection described in Eq.~(\ref{decoupled}) leads to an important consequence that the parity of each symmetry sector, defined by $Q_{\alpha}=\omega^{(N+1)/2}\alpha_{1}\alpha_{2}^\dagger$ and $Q_{\beta}=\omega^{(N+1)/2}\beta_{1}\beta_{2}^\dagger$, should be conserved. To see this, we consider the generic non-local interacting terms allowed in the PZMs that
\begin{eqnarray}
H_I=\sum_{k,l,m,n}\xi_{k,l,m,n}\alpha_{1}^{k}\alpha_{2}^{l}\beta_{1}^{m}\beta_{2}^{n}+h.c.,
\end{eqnarray}
where $k+l+m+n=0\text{ mod }N$ due to the total parity conservation. The symmetry protection condition in Eq.~(\ref{decoupled}) further requires that $k+l=m+n=0\text{ mod }N$. Then the interaction term reads $H_I=\sum_{k,m}\xi_{k,m}Q_{\alpha}^{k}Q_{\beta}^{m}+h.c.$, which commutes with the parity operator of each sector. Note that the interaction term couples $\alpha_{i}$-sector and $\beta_{i}$-sector when $k \neq 0$, $m \neq 0$, and the parity conservation of each sector does not imply the decomposition of the whole interacting system into decoupled subsystems.

\subsection{Effective braiding Hamiltonian}

\begin{figure}
\centering
\includegraphics[width=\columnwidth]{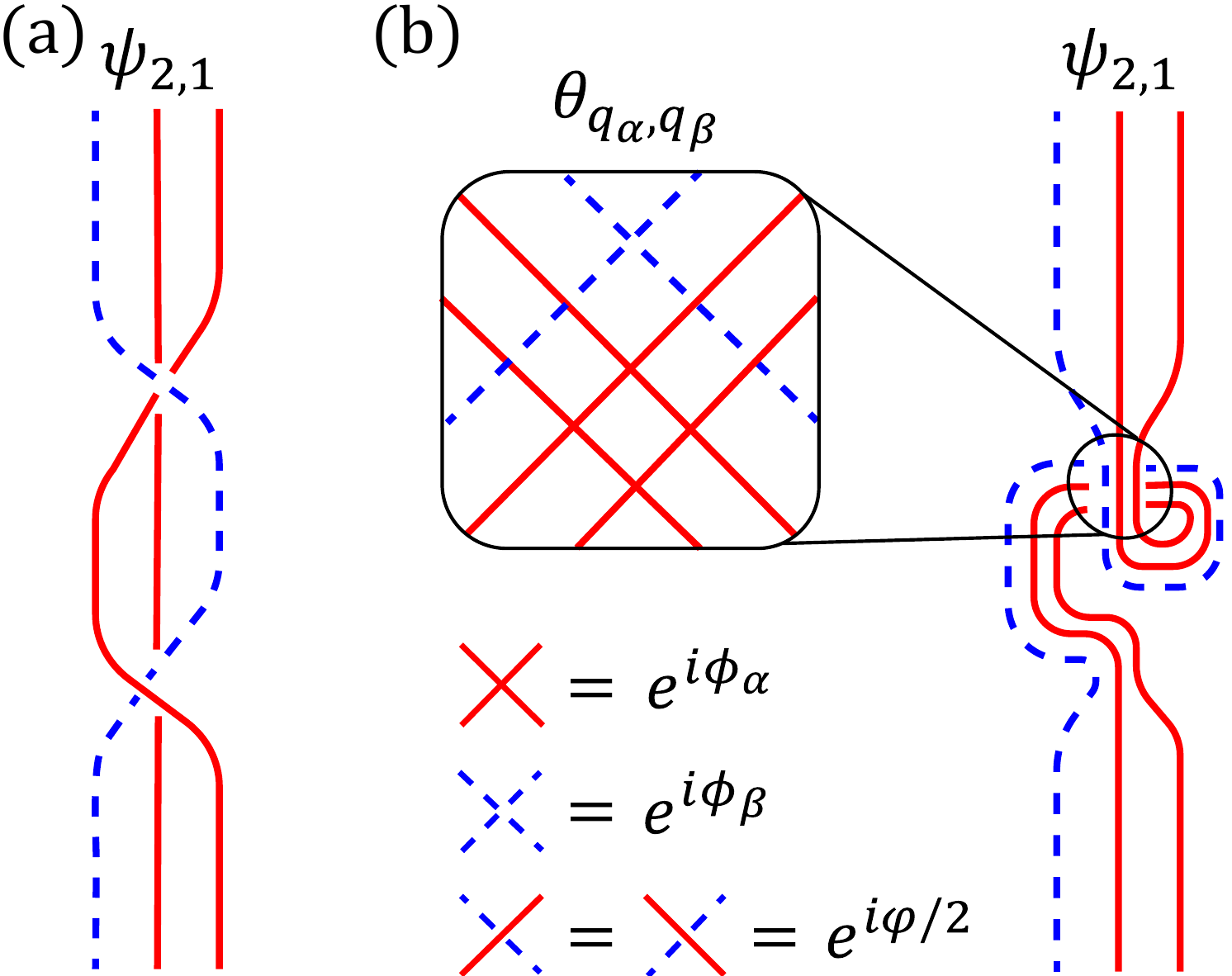}
\caption{Topological spin of Abelian anyon $\psi_{2,1}$. (a) $\psi_{2,1}$ is a composite of two anyons of the type $\psi_{1,0}$ (denoted by the red solid lines) and one anyon of the type $\psi_{0,1}$ (denoted by the blue dashed lines). The topological spin is defined as the phase caused by $2\pi$ rotation of this composite object relative to the rest of the system. The rotation in (a) is topologically equivalent to the multiple crossings in (b). The red-red, blue-blue and red-blue crossings give phase factors of $\phi_{\alpha}$, $\phi_{\beta}$ and $\varphi/2$, thus we obtain the topological spin $\theta_{2,1}=\exp[i(4\phi_{\alpha}+\phi_{\beta}+2\varphi)]$. More generally,  the topological spin of anyon $\psi_{q_{\alpha},q_{\beta}}$ is $\theta_{q_{\alpha},q_{\beta}}=\exp[i(q_{\alpha}^{2}\phi_{\alpha}+q_{\beta}^{2}\phi_{\beta}+q_{\alpha}q_{\beta}\varphi)]$.}
\label{topospin}
\end{figure}

We now turn to studying the braiding of PZMs through an effective braiding Hamiltonian under the unitary symmetry protection. By definition, the braiding process is an adiabatic evolution $U(T)=\mathcal{T}\exp\bigr[{-i\int_{-T/2}^{T/2}H(\tau)d\tau}\bigr]$, with the Hamiltonian returning to initial one after braiding $H(-T/2)=H(T/2)$ and $\mathcal{T}$ the time-ordering operator, and can be described by an effective Floquet Hamiltonian $H_E=(i/T)\ln{U(T)}$. If the unitary symmetry preserves instantaneously, such that $[H(\tau),S]=0$ for every moment within the interval $-T/2<\tau<T/2$, then the effective Hamiltonian also satisfies the symmetry $[H_E,S]=0$ as the unitary time evolution operator $U(T)$ commutes with unitary symmetry. Consequently, the effective Hamiltonian preserves the parity of each sector, and it must take the generic form as
\begin{equation}
H_{E}=\sum_{m=1}^{N-1} a_{m}Q_{\alpha}^{m}+\sum_{n=1}^{N-1} b_{n}Q_{\beta}^{n}+\sum_{k,l=1}^{N-1} c_{k,l}Q_{\alpha}^{k}Q_{\beta}^{l},\label{effHam}
\end{equation}
where $a_{m}^{*}=a_{N-m}$, $b_{n}^{*}=b_{N-n}$, $c_{k,l}^{*}=c_{N-k,N-l}$
for hermitianity. The first and second terms in Eq.~(\ref{effHam}) denote the exchange Hamiltonian of two sectors. The third term in $H_E$ couples $\alpha_{i}$-sector and $\beta_{i}$-sector and indicates that, with only parity conservation, time-periodic dynamics involving two pairs of PZMs cannot be divided into two independent processes. Thus to prove the SPNA statistics, we shall show that the third term in Eq.~(\ref{effHam}) should indeed be absent in the braiding.

The braiding of two pairs of PZMs should satisfy anyon theory and spin-statistics theorem. The parity conservation of each sector leads to the following fusion rules of the unitary symmetry-protected PZMs
\begin{eqnarray}
F\times F & = & \sum_{q_{\alpha},q_{\beta}=0}^{N-1}\psi_{q_{\alpha},q_{\beta}},\label{fusion1}\\
\psi_{q_{\alpha},q_{\beta}}\times\psi_{q'_{\alpha},q'_{\beta}} & = & \psi_{q_{\alpha} \oplus q'_{\alpha},q_{\beta} \oplus q'_{\beta}},\label{fusion}
\end{eqnarray}
where $a \oplus b = (a + b) \text{ mod }N$, $F$ denotes a domain wall with a pair of PZMs, or equivalently, a single zero-energy Fock parafermion mode~\cite{FockPara} and $\psi_{q_{\alpha},q_{\beta}}$ denotes the Abelian anyon with $q_{\alpha(\beta)}$ quasiparticles of $\alpha_{i}(\beta_{i})$-sector. Since the Abelian anyon $\psi_{q_{\alpha},q_{\beta}}$ is a composite object, its topological spin $\theta_{q_{\alpha},q_{\beta}}$ can be deduced from those of $\psi_{1,0}$, $\psi_{0,1}$ and $\psi_{1,1}$. Suppose that the topological spins of $\psi_{1,0}$, $\psi_{0,1}$, $\psi_{1,1}$ are $\theta_{1,0}=\exp(i\phi_{\alpha})$, $\theta_{0,1}=\exp(i\phi_{\beta})$, $\theta_{1,1}=\exp[i(\phi_{\alpha}+\phi_{\beta}+\varphi)]$ separately. Then the general expression for topological spin is $\theta_{q_{\alpha},q_{\beta}}=\exp[i(q_{\alpha}^{2}\phi_{\alpha}+q_{\beta}^{2}\phi_{\beta}+q_{\alpha}q_{\beta}\varphi)]$ as shown in Fig.~\ref{topospin} where $\theta_{2,1}$ is depicted as an example. The $\mathbb{Z}_{N}\times\mathbb{Z}_{N}$ structure of $\theta_{q_{\alpha},q_{\beta}}$ enforced by the fusion rules further requires $\theta_{q_{\alpha}+N,q_{\beta}}=\theta_{q_{\alpha},q_{\beta}}$ and $\theta_{q_{\alpha},q_{\beta}+N}=\theta_{q_{\alpha},q_{\beta}}$, which give
\begin{equation}
\theta_{q_{\alpha},q_{\beta}}=\exp\left[i\frac{\pi}{N}\left(q_{\alpha}^{2}u+q_{\beta}^{2}v+2q_{\alpha}q_{\beta}w\right)\right],\label{thetaqq}
\end{equation}
where $u$, $v$, $w$ are integer numbers. Label the eigenstate of the system $|q_{\alpha},q_{\beta}\rangle$ by the presence of Abelian anyon $\psi_{q_{\alpha},q_{\beta}}$, in which the braiding $U(T)$ takes a matrix form with only diagonal matrix elements $\left[U(T)\right]_{q_{\alpha},q_{\beta}}=\langle q_{\alpha},q_{\beta}|U(T)|q_{\alpha},q_{\beta}\rangle$. From the spin-statistics theorem $\theta_{q_{\alpha},q_{\beta}}=\theta_{F}^{2}\left[U(T)\right]_{q_{\alpha},q_{\beta}}^{2}$~\cite{Kitaev2006}, where $\theta_F$ denotes the topological spin of a Fock parafermion mode, we find
\begin{eqnarray}
\left[U(T)\right]_{q_{\alpha},q_{\beta}} = (-1)^{f(q_\alpha,q_\beta)} e^{i\pi(q_{\alpha}^{2}u+q_{\beta}^{2}v+2q_{\alpha}q_{\beta}w)/2N}.\label{braiddiag}
\end{eqnarray}
Here ${f(q_\alpha,q_\beta)}$ is an integer-valued function. This formula tells that the elements of $U(T)$ can only take discrete values and $\left[U(T)\right]^{4N}=1$. From the Eq~(\ref{braiddiag}), we can further obtain the diagonal matrix element of the effective braiding Hamiltonian $\langle q_{\alpha},q_{\beta}|H_E|q_{\alpha},q_{\beta}\rangle = (q_{\alpha}^{2}u+q_{\beta}^{2}v+2q_{\alpha}q_{\beta}w)[\pi/(2NT)] + (\pi/T) f(q_\alpha,q_\beta)$, which determines that the coefficients $a_m$, $b_n$, $c_{k,l}$ in Eq.~(\ref{effHam}) must also take discrete values.

The symmetry of the braiding Hamiltonian in Eq.~(\ref{effHam}) and the discreteness of $U(T)$ suffice to prove the unitary SPNA statistics of PZMs. The discreteness of the matrix elements in $U(T)$ implies that the coefficients $(a_m, b_n, c_{k,l})$ in the braiding Hamiltonian cannot change continuously. This key feature facilitates to determine the coefficients from a special case that the whole system hosting PZMs is formed by two independent subsystem. The generic system can be obtained from such special case by adiabatically switching on couplings and interactions between the subsystems without undergoing topological phase transition. Then the braiding matrix obtained in the special case must be applicable to the generic case. In a special case that the PZMs $\alpha_{1,2}$ and $\beta_{1,2}$ belong to two fully decoupled subsystems, we must have that $c_{k,l}=0$ for all $k$ and $l$. 
With this we conclude that the generic braiding operator reads
\begin{equation}
U(T)=\frac{1}{N}\left(\sum_{m=0}^{N-1}\lambda_{\alpha,m}Q_{\alpha}^{m}\right)\otimes\left(\sum_{n=0}^{N-1}\lambda_{\beta,n}Q_{\beta}^{n}\right),
\label{UnitaryParaBraid}\end{equation}
where $\lambda_{\alpha(\beta),m}=\omega^{m(m+N+2r_{\alpha(\beta)})/2}$ with $r_{\alpha(\beta)}=0,\cdots,N-1$. The above result shows that the braiding of the PZM pairs with unitary symmetry protection can always be reduced to braiding the two copies of PZMs individually, even though the whole system cannot be characterized by two decoupled subsystems due to the generally existing strong interactions. This braiding operation is highly nonlinear for $N>2$, sharply contrast to the Majorana counterparts with symmetry protection~\cite{XJL2014PRX,Gao2016,JSH2022}, in which the braiding transformation is linear.

\section{Braiding matrix in a tri-junction}\label{SecJunction}

\begin{figure*}
\centering
\includegraphics[width=\textwidth]{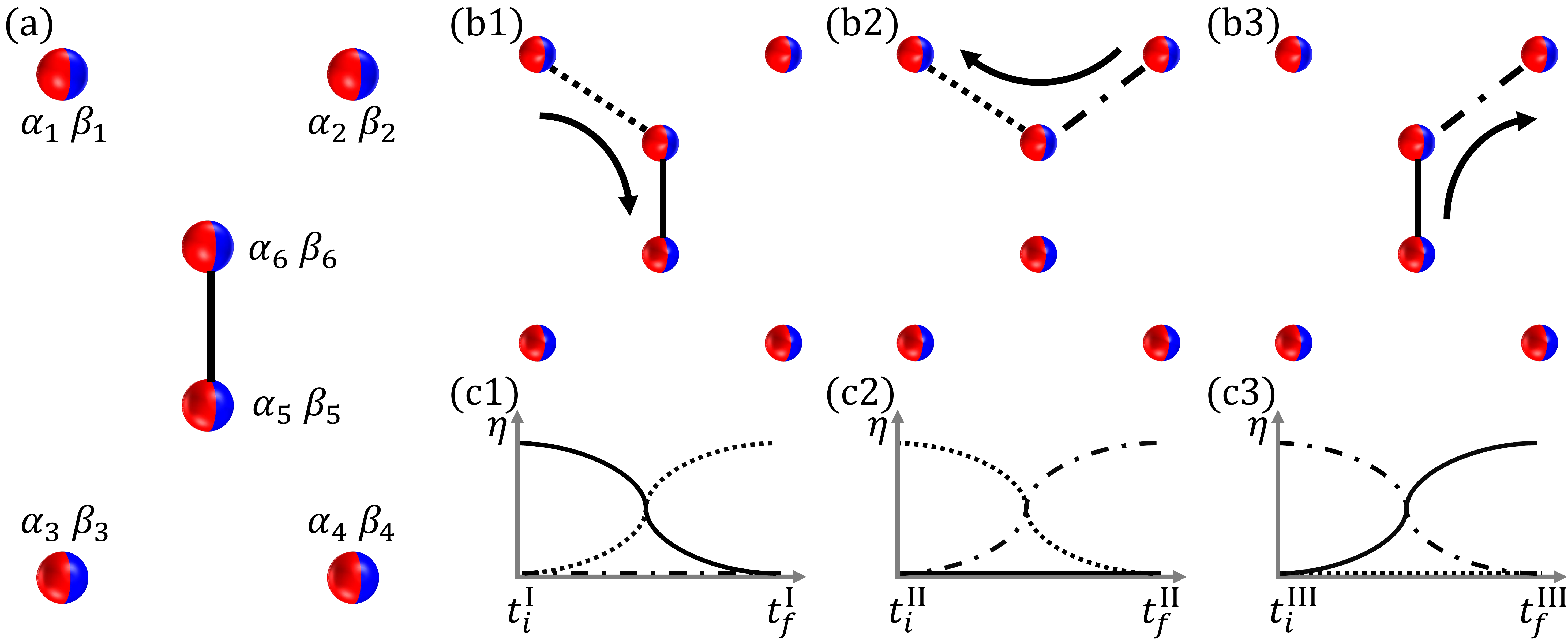}
\caption{Braiding process in a tri-junction. (a) Four pairs of PZMs $\alpha_{1,\cdots,4}$ and $\beta_{1,\cdots,4}$ constitude a two-$N$-qudit system (with definite total parity of each sector) and $\alpha_{5,6}$, $\beta_{5,6}$ serve as ancilla modes to form a tri-junction and braid PZMs in the qudits. The system is initialized by coupling $\alpha_{5,6}$, $\beta_{5,6}$ together. (b1-b3) During the braiding of $\alpha_1(\beta_1)$ and $\alpha_2(\beta_2)$, the PZMs $\alpha_{3,4}$, $\beta_{3,4}$ are decoupled from the rest PZMs. The ancilla modes $\alpha_6(\beta_6)$ are coupled with $\alpha_5(\beta_5)$, $\alpha_1(\beta_1)$, $\alpha_2(\beta_2)$, $\alpha_5(\beta_5)$ in succession. The PZMs are effectively transported according to the arrows. The groundstate degeneracy is kept as $N^2$-fold in the process. (c1-c3) Time dependence of the coupling strength. $t_i^{\text{II}}=t_f^{\text{I}}$ and $t_i^{\text{III}}=t_f^{\text{II}}$. From $t_i^{\text{I}}$ to $t_f^{\text{III}}$, the Hamiltonian is changed back to its original form while the PZMs $\alpha_1$, $\beta_1$ and $\alpha_2$, $\beta_2$ are exchanged adiabatically.}
\label{junction}
\end{figure*}

To further confirm the SPNA statistics of PZMs, we study the braiding dynamics by explicitly deriving the non-Abelian geometric phases resulting from adiabatic manipulation of PZMs within a tri-junction configuration in Fig.~\ref{junction}. The exchange of PZMs is accomplished through a three-point turn in the tri-junction, with the groundstate degeneracy remaining invariant throughout the adiabatic exchange process. In the following subsections, we expound upon the braiding dynamics within the tri-junction and delve into the characteristics of degenerate groundstates of the system. Subsequently, we derive the recursion relation for non-Abelian Berry phases accumulated during the braiding process. Finally, we show that the Berry phases necessarily adopt a symmetric form, as dictated by the adiabatic condition.

\subsection{Tri-junction and groundstate subspace}

The braiding process consists of three subsequent stages of the adiabatic time evolution. Consider six pairs of symmetry-protected PZMs $\alpha_{1,\cdots,6}$ and $\beta_{1,\cdots,6}$ as shown in Fig.~\ref{junction}(a) and suppose that we are going to braid $\alpha_{1}$, $\beta_{1}$ and $\alpha_{2}$, $\beta_{2}$. Before the braiding, we initialize the system by coupling $\alpha_{5}$, $\beta_{5}$ and $\alpha_{6}$, $\beta_{6}$ together, then only $\alpha_{1,\cdots,4}$ and $\beta_{1,\cdots,4}$ remain zero-energy. Since the total parity for each sector is conserved for symmetry-protected parafermionic systems during the braiding, if the parity for $\alpha(\beta)$-sector is fixed, then the groundstates after initialization of this system is $N \times N$-fold degenerate. At the first stage of braiding process, the coupling between $\alpha_{1}$, $\beta_{1}$ and $\alpha_{6}$, $\beta_{6}$ is turned on adiabatically, and the coupling between $\alpha_{5}$, $\beta_{5}$ and $\alpha_{6}$, $\beta_{6}$ is turned off adiabatically at the same time. The coupling Hamiltonian for this stage is written as $H_{\text{I}}(t)=\eta_{16}(t)H_{16}+\eta_{56}(t)H_{56}$, where $\eta_{16}(t)$, $\eta_{56}(t)$ are depicted in Fig.~\ref{junction}(c1) and $H_{ij}$ denotes the coupling term between $\alpha_{i}$, $\beta_{i}$ and $\alpha_{j}$, $\beta_{j}$, whose concrete form will be specified later. The second and third stages of the braiding process are governing by similar Hamiltonians $H_{\text{II}}(t)=\eta_{26}(t)H_{26}+\eta_{16}(t)H_{16}$ and $H_{\text{III}}(t)=\eta_{56}(t)H_{56}+\eta_{26}(t)H_{26}$ as depicted in Fig~\ref{junction}. After all the three stages, the Hamiltonian of the whole system returns to the original form after initialization and this braiding process is completed. During the braiding, there are $N \times N$ degenerate groundstates, and since $\alpha_{3}$, $\beta_{3}$ and $\alpha_{4}$, $\beta_{4}$ are zero-modes at each stage, the instantaneous goundstates can always be labelled as the parity eigenstates of $\alpha_{3}$, $\alpha_{4}$ and $\beta_{3}$, $\beta_{4}$
\begin{eqnarray}
Q_{X,34}|\psi(q'_{\alpha},q'_{\beta},t)\rangle=\omega^{q'_{X}}|\psi(q'_{\alpha},q'_{\beta},t)\rangle,
\end{eqnarray}
where $Q_{X,34}=\omega^{(N+1)/2}X_{3}X_{4}^{\dagger}$ with $X=\alpha,\beta$, and $q'_{X}$ is $q_{X,34}$ for short. In these bases, the effect of braiding is manifested in the Berry phase of the groundstate $|\psi(q'_{\alpha},q'_{\beta},t_f)\rangle=e^{i\chi(q'_{\alpha},q'_{\beta})}|\psi(q'_{\alpha},q'_{\beta},t_i)\rangle$, where $t_{i(f)}$ denotes the initial (final) time of the braiding dynamics.

\subsection{Recursion relation for Berry phases}

Since the braiding matrix is diagonal in the bases of the eigenstates of $Q_{\alpha,34}$ and $Q_{\beta,34}$, we can obtain the matrix by deriving the recursion relation for Berry phases $\chi(q'_{\alpha},q'_{\beta})$ of the groundstates $|\psi(q'_{\alpha},q'_{\beta},t)\rangle$ accumulated during the braiding. For each stage, we can identify several symmetry operators that commute with the time evolution operator and transform a groundstate to another. Then the corresponding Berry phases are also related by the symmetry operators. Take the first stage as an example. The PZMs $\alpha_{2,3,4}$ and $\beta_{2,3,4}$ are not involved in the dynamics, then the parity operators $Q_{\alpha,23}$ and $Q_{\beta,23}$ are suitable symmetry operators since they commute with the time evolution operator of this stage $U_{\text{I}}=\mathcal{T}\exp[-i\int_{t^{\text{I}}_{i}}^{t^{\text{I}}_{f}} H_{\text{I}}(t)dt]$, with $t^{\text{I}}_{i(f)}$ denoting the initial (final) time of the first stage, and can advance the quantum numbers of groundstates by one increment as
\begin{eqnarray}
Q_{\alpha,23}|\psi(q'_{\alpha},q'_{\beta},t)\rangle&=&e^{i\delta_{\alpha}^{\text{I}}}|\psi(q'_{\alpha}+1,q'_{\beta},t)\rangle,\\
Q_{\beta,23}|\psi(q'_{\alpha},q'_{\beta},t)\rangle&=&e^{i\delta_{\beta}^{\text{I}}}|\psi(q'_{\alpha},q'_{\beta}+1,t)\rangle,
\end{eqnarray}
where $\delta_{\alpha(\beta)}^{\text{I}}$ is short for $\delta_{\alpha(\beta)}^{\text{I}}(q'_{\alpha},q'_{\beta},t)$. And we have $U_{\text{I}}Q_{X,23}|\psi(t^{\text{I}}_i)\rangle=e^{i\chi_{\text{I}}}Q_{X,23}|\psi(t^{\text{I}}_f)\rangle$ with $X=\alpha$, $\beta$ since $U_{\text{I}}$ and $Q_{X,23}$ commute. It follows that $\chi_{\text{I}}(q'_{X}+1)=\chi_{\text{I}}(q'_{X})+\delta_{X}^{\text{I}}(t_f^{\text{I}})-\delta_{X}^{\text{I}}(t_i^{\text{I}})$. Similar identities can be obtained for the second (third) stage by replacing the symmetry operators $Q_{X,23}$ by $Q_{X,35(13)}$. By defining $\delta_{X,i(f)}=\delta_{X}^{\text{I}}(t_{i(f)}^{\text{I}})+\delta_{X}^{\text{II}}(t_{i(f)}^{\text{II}})+\delta_{X}^{\text{III}}(t_{i(f)}^{\text{III}})$, we obtain
\begin{eqnarray}
\chi(q'_{\alpha}+1,q'_{\beta})&=&\chi(q'_{\alpha},q'_{\beta})+\delta_{\alpha,f}-\delta_{\alpha,i},\\
\chi(q'_{\alpha},q'_{\beta}+1)&=&\chi(q'_{\alpha},q'_{\beta})+\delta_{\beta,f}-\delta_{\beta,i}.
\end{eqnarray}
From the above recursion relations, we see that if the phase factor $\delta_{X,i(f)}(q'_{\alpha},q'_{\beta})$ can be divided into two independent parts $\delta_{X,i(f)}(q'_{\alpha},q'_{\beta})=\delta^{\alpha}_{X,i(f)}(q'_{\alpha})+\delta^{\beta}_{X,i(f)}(q'_{\beta})$, where $\delta^{\alpha}_{X,i(f)}(\delta^{\beta}_{X,i(f)})$ only relies on the parity $q'_\alpha (q'_\beta)$, the total Berry phase $\chi(q'_{\alpha},q'_{\beta})$ can be similarly decomposed.

\subsection{Decoupled braiding matrix}

The adiabatic condition determines the decomposition of the phase factor $\delta_{X,i(f)}$ and the non-Abelian Berry phases. To see this, we explore the concrete form of the coupling Hamiltonian $H_{ij}$. Most generically,
\begin{equation}
H_{ij}=H_{ij}^{\alpha}+H_{ij}^{\beta}+H_{ij}^{\text{int}},
\end{equation}
where $H_{ij}^{\alpha}=\sum_{m=1}^{N-1} a'_{m}Q_{\alpha,ij}^{m}$, $H_{ij}^{\beta}=\sum_{n=1}^{N-1} b'_{n}Q_{\beta,ij}^{n}$ and $H_{ij}^{\text{int}}=\sum_{k,l=1}^{N-1} c'_{k,l}Q_{\alpha,ij}^{k}Q_{\beta,ij}^{l}$. The PZMs of different sectors can be coupled together by $H_{ij}^{\text{int}}$. At the initial and final moments of each stage $t_{i(f)}^J$, $J=\text{I, II, III}$, the time-dependent Hamiltonian $H_{J}(t)$ only involves a single coupling Hamiltonian $H_{ij}$. According to the adiabatic condition, except for the degenerate states of the PZMs, any extra ground degeneracy should be avoided in the braiding. Thus the groundstate of $H_{ij}$ must be non-degenerate. Then the $N^2$-fold groundstates of the whole system at time $t_{i(f)}^J$ must be eigenstates of the involved parity operators $Q_{\alpha,ij}$ and $Q_{\beta,ij}$. For example, at the initial and final moments of the first stage
\begin{eqnarray}
Q_{X,56}|\psi(q'_{\alpha},q'_{\beta},t_{i}^\text{I})\rangle&=&\omega^{q_{X,56}}|\psi(q'_{\alpha},q'_{\beta},t_{i}^\text{I})\rangle,\label{measurement1}\\
Q_{X,16}|\psi(q'_{\alpha},q'_{\beta},t_{f}^\text{I})\rangle&=&\omega^{q_{X,16}}|\psi(q'_{\alpha},q'_{\beta},t_{f}^\text{I})\rangle.\label{measurement2}
\end{eqnarray}
Certain Hamiltonians like $H_{ij}^{\alpha}=H_{ij}^{\beta}=0$, $H_{ij}^{\text{int}}=c'_{1,1}Q_{\alpha,ij}Q_{\beta,ij}+h.c.$ are forbidden since their groundstates are degenerate. Then the physical coupling Hamiltonian must be continuously connected to the simple one $H'_{ij} = a'_{1}Q_{\alpha,ij} + b'_{1}Q_{\beta,ij} +h.c.$, whose groundstate is non-degenerate. The Hamiltonian $H'_{ij}$ takes a decoupled form and, as a result, the braiding matrix can be divided into two independent matrices for the two sectors of PZMs with the matrix elements of the Berry phases
\begin{equation}
\chi(q'_{\alpha},q'_{\beta})=\frac{\pi}{N}\left[(q'_{\alpha}-k_{\alpha})^{2}+(q'_{\beta}-k_{\beta})^{2}\right],
\end{equation}
where $k_{\alpha(\beta)}$ is integer. This result can also be obtained by effectively regarding the braiding process as successive forced measurements~\cite{Bonderson2008}. The Eqs.~(\ref{measurement1}) and (\ref{measurement2}) indicate that during each stage, the parity of each sector can be measured independently, and thus the braiding phases can be divided into two independent parts. Switch the focus from parity eigenstates to zero-mode operators, we find the non-Abelian Berry phases $\chi(q'_{\alpha},q'_{\beta})$ correspond to the following transformation of PZM operators
\begin{eqnarray}
\begin{split}
\alpha_1 \to s_\alpha \alpha_2,&&\alpha_2 \to \omega s_\alpha \alpha_1^\dagger \alpha_2^2,\\
\beta_1 \to s_\beta \beta_2,&&\beta_2 \to \omega s_\beta \beta_1^\dagger \beta_2^2,
\end{split}
\end{eqnarray}
where $s_{\alpha(\beta)}=\omega^{k_{\alpha(\beta)}+N/2}$. This transformation is also consistent with Eq.~(\ref{UnitaryParaBraid}) with $k_{\alpha(\beta)}=-r_{\alpha(\beta)}-N/2$. The above result shows that each PZM in one pair sees only one PZM rather than both parafermion modes in another pair during the braiding process, manifesting a central feature of the SPNA statistics for parafermions.

\section{Symmetry-protected PZMs in interacting quantum wires}\label{SecModel}

Finally we propose a physical system based on strongly correlated nanowires to realize a pair of PZMs at each edge protected by mirror symmetry. In this model, the PZMs are naturally classified into two symmetry sectors with conserved parity. Further, symmetry-preserving interactions can generally mix the PZMs of the two sectors without ruining the non-Abelian statistics, which serves as a non-trivial demonstration of the generic theory.

\begin{figure}
\centering
\includegraphics[width=\columnwidth]{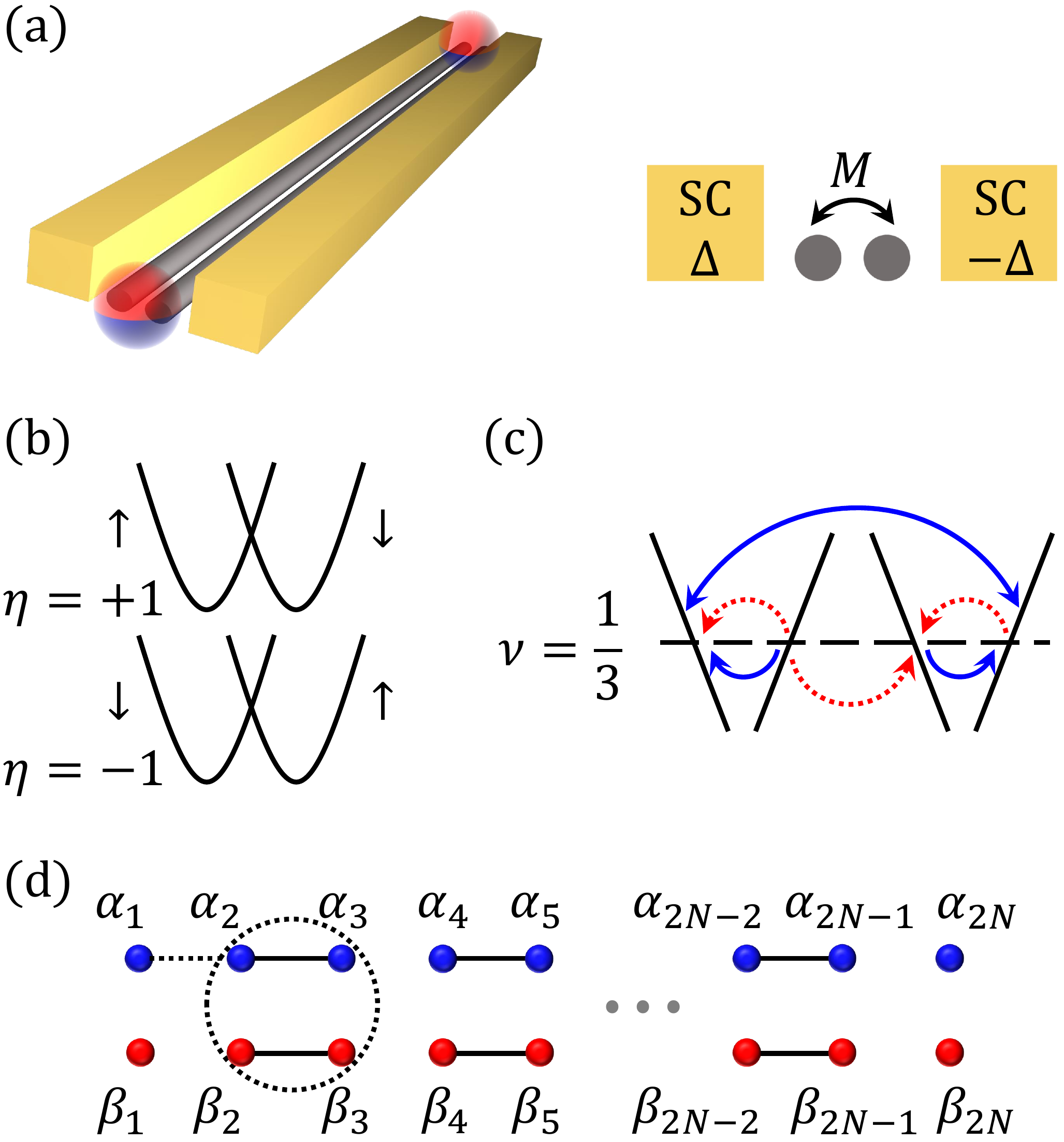}
\caption{Mechanism for the mirror symmetric model and local mixing under perturbation. (a) Two spin-orbit coupled nanowires with opposite spin-orbit couplings are placed in between two conventional superconductors with $\pi$-phase difference. The nanowires are coupled by tunneling $M$. Pairs of PZMs are localized at the ends of the system. (b) The band structure of the nanowires when $M=\Delta=0$. (c) At fractional filling $\nu=1/3$, pairing terms (blue solid lines) and backscattering terms (red dashed lines) assisted by interactions are dominant, giving rise to $\mathbb{Z}_6$ PZMs. (d) Dimerized lattice model hosting symmetry-protected PZMs. The dashed line and circle denote the symmetric perturbation.}
\label{Model}
\end{figure}

\subsection{Model with mirror symmetry}

We start with a non-interacting model hosting pairs of symmetry-protected MZMs and then turn to the strongly-correlated regime. As depicted in Fig.~\ref{Model}(a), the model consists of two spin-orbit coupled nanowires proximited to two conventional superconductors with $\pi$-phase difference and the Bloch Hamiltonian reads
\begin{equation}
\mathcal{H}(k)= (\varepsilon_k -\mu) \tau_z-\alpha k \eta_z \sigma_z+M\eta_x \tau_z+\Delta \eta_z \tau_y \sigma_y,
\end{equation}
where $\eta_i$, $\tau_i$, $\sigma_i$ are Pauli matrices acting in left/right-side, particle-hole and spin subspaces. The spin-orbit couplings of the two nanowires are opposite and in the absence of superconductivity $\Delta$ and inter-wire tunneling $M$, the band structure is as shown in Fig.~\ref{Model}(b). This system respects time-reversal symmetry $\mathcal{T}\mathcal{H}(k)\mathcal{T}^{-1}=\mathcal{H}(-k)$ and particle-hole symmetry $\mathcal{P}\mathcal{H}(k)\mathcal{P}^{-1}=-\mathcal{H}(-k)$ with $\mathcal{T}=i\sigma_y \mathcal{K}$ and $\mathcal{P}=\tau_x \mathcal{K}$, and thus belongs to the symmetry class DIII. The system is also invariant under a unitary mirror symmetry $\mathcal{M}\mathcal{H}(k)\mathcal{M}^{-1}=\mathcal{H}(k)$ with $\mathcal{M}=\eta_x \sigma_x$, and we can block-diagonalize the Hamiltonian $\mathcal{H}(k)$ with respect to the two eigenvalues of $\mathcal{M}$. In the topological regime, both blocks host MZMs as edge states. The two MZMs are eigenstates of different eigenvalues of the symmetry operator $\mathcal{M} \gamma \mathcal{M}^{-1}=\gamma$, $\mathcal{M} \bar{\gamma} \mathcal{M}^{-1}=-\bar{\gamma}$, and thus are protected by the mirror symmetry $\mathcal{M} i\gamma\bar{\gamma} \mathcal{M}^{-1}=-i\gamma\bar{\gamma}$. To realize symmetry-protected PZMs beyond MZMs, we tune the chemical potential to fractional filling $\nu=1/(2n+1)$, and introduce non-commutative interaction terms to open different gaps. These gaps effectively form domain walls at the edges of the system, leading to the appearance of PZMs. This physical picture is made clear by adopting the following bosonization formalism. The spectra around Fermi points are linearized as $\psi_{\eta'\sigma'}=R_{\eta'\sigma'}e^{ik_F^{1\eta'\sigma'}x}+L_{\eta'\sigma'}e^{ik_F^{\bar{1}\eta'\sigma'}x}$, where $\psi_{\eta'\sigma'}$ denote the bases after block-diagonalization $\mathcal{M}\psi_{\eta'\sigma'}\mathcal{M}^{-1}=\eta'\psi_{\eta'\sigma'}$ and $\eta'=\pm 1$, $\sigma'=\pm 1$. The non-commutative backscattering and pairing terms that open different gaps are written as
\begin{align}
H_M&=\tilde{M}\sum_{\eta'}[R^{\dagger}_{\eta'\bar{\eta}'}L_{\eta'\bar{\eta}'}]^{n}
R^{\dagger}_{\eta'\bar{\eta}'}L_{\eta'\eta'}
[R^{\dagger}_{\eta'\eta'}L_{\eta'\eta'}]^{n},\\
H_{\Delta}&=\tilde{\Delta}\sum_{\eta'}[R^{\dagger}_{\eta'\eta'}L_{\eta'\eta'}]^{n}
R^{\dagger}_{\eta'\eta'}L^{\dagger}_{\eta'\bar{\eta}'}
[R_{\eta'\bar{\eta}'}L^{\dagger}_{\eta'\bar{\eta}'}]^{n},
\end{align}
where $\bar{\eta}'$ is short for $-\eta'$. In the spetical case $n=1$, the filling is $\nu=1/3$, and the processes that constitude $H_M$ and $H_{\Delta}$ are depicted in Fig.~\ref{Model}(c). These terms can be rewritten as $H_M=2\tilde{M}\sum_{\eta'}\cos[\varphi_{1\eta'\bar{\eta}'}-\varphi_{\bar{1}\eta'\eta'}]$ and $H_{\Delta}=2\tilde{\Delta}\sum_{\eta'}\cos[\varphi_{1\eta'\eta'}+\varphi_{\bar{1}\eta'\bar{\eta}'}]$ after bosonization procedure with $\varphi_{r\eta'\sigma'}=(n+1)\phi_{r\eta'\sigma'}-n\phi_{\bar{r}\eta'\sigma'}$ and $R_{\eta'\sigma'}=e^{i\phi_{1\eta'\sigma'}}$, $L_{\eta'\sigma'}=e^{i\phi_{\bar{1}\eta'\sigma'}}$. By applying the unfolding method~\cite{Stern2014,Giamarchi}, the gaps opened by the two terms form a domain wall at edge, where we identify a pair of $\mathbb{Z}_{2(2n+1)}$ PZMs
\begin{eqnarray}
\begin{split}
\alpha&=&e^{i[(\varphi_{11\bar{1}}-\varphi_{\bar{1}11})+(\varphi_{111}+\varphi_{\bar{1}1\bar{1}})]/[2(2n+1)]},\\
\beta&=&e^{i[(\varphi_{1\bar{1}1}-\varphi_{\bar{1}\bar{1}\bar{1}})+(\varphi_{1\bar{1}\bar{1}}+\varphi_{\bar{1}\bar{1}1})]/[2(2n+1)]}.
\end{split}
\end{eqnarray}
These PZMs are protected by the mirror symmetry since $\mathcal{M}\alpha\mathcal{M}^{-1}=\alpha$ and $\mathcal{M}\beta\mathcal{M}^{-1}=\omega\beta$, satisfying the generic theory.  Since we have defined the PZMs in the eigenbasis of the mirror symmetry, they are naturally classified into two symmetry sectors with conserved parity.

\subsection{Effect of nonlinear interactions}

As mentioned in the generic theory, the parity conservation of each sector does not imply the decomposition of the whole system into two decoupled subsystems. To show that certain interactions can mix modes in two sectors without ruining non-Abelian statistics, we model the above system with a lattice Hamiltonian (see Fig.~\ref{Model}(d))
\begin{equation}
H_{\text{latt}}=H_{\alpha}+H_{\beta}+H_{p},\label{latticeH}
\end{equation}
where $H_{X}=\sum_{i=1}^{N-1}JX_{2i}^\dagger X_{2i+1}+J^* X_{2i+1}^\dagger X_{2i}$ and $\alpha_i$, $\beta_i$ are parafermionic operators that satisfy $\mathcal{M}\alpha_i\mathcal{M}^{-1}=\alpha_i$, $\mathcal{M}\beta_i\mathcal{M}^{-1}=\omega\beta_i$. The third part $H_p$ includes certain symmetric perturbation terms $H_p=f\alpha_1^\dagger \alpha_2+f^* \alpha_2^\dagger \alpha_1+V\alpha_2^\dagger \alpha_3 \beta_2^\dagger \beta_3 +V^* \beta_3^\dagger \beta_2 \alpha_3^\dagger \alpha_2$, where $f/J\ll1$, $V/J\ll1$, $f/V\sim\mathcal{O}(1)$ and $[H_p,\mathcal{M}]=0$. In the absence of $H_p$, the lattice system is in a dimerized phase, with $\alpha_{1,2N}$, $\beta_{1,2N}$ being exact parafermion zero edge modes, which characterizes the results in the above nanowire model. The perturbation terms in $H_p$ are only introduced to the zero mode $\alpha_1$ and the new PZM in the presence of $H_p$ can be constructed as $\bar{\alpha}_{1}=\alpha_{1}+\sum_{i}m_{i}\hat{O}_{i}^{(1)}+\sum_{j}n_{j}\hat{O}_{j}^{(2)}+\cdots$, where $m_{i}\sim\mathcal{O}(f/J)$ and $n_{j}\sim\mathcal{O}(f^{2}/J^{2})$. By calculating the perturbation series order by order, we find $\tilde{\alpha}_{1}$ cannot be constructed solely by terms with $\alpha_{i}$ , and its dynamics is governed by the full Hamiltonian rather than just $H_{\alpha}$ (see Appendix~\ref{appenC}). For an intuitive result, if $\alpha_{i}$ and $\beta_{i}$ are reduced to Majorana operators $\gamma_{i}^{\alpha}$ and $\gamma_{i}^{\beta}$, the exact zero mode can be constructed explicitly
\begin{equation}
\bar{\gamma}_{1}^{\alpha}=\gamma_{1}^{\alpha}+a\gamma_{3}^{\alpha}+ib\gamma_{2}^{\beta}\gamma_{3}^{\alpha}\gamma_{3}^{\beta}+ic\gamma_{1}^{\alpha}\gamma_{2}^{\beta}\gamma_{3}^{\beta},
\end{equation}
where $a\sim\mathcal{O}(f/J)$, $b\sim\mathcal{O}(f^2/J^2)$ and $c\sim\mathcal{O}(f^3/J^3)$. From this expression, we see that the Majorana modes $\beta_{2,3}$ in $H_{\beta}$ are involved in the wavefunction of $\bar{\gamma}_1^\alpha$, which is an MZM in $\alpha$-sector since $\mathcal{M}\bar{\gamma}_1^\alpha\mathcal{M}^{-1}=\bar{\gamma}_1^\alpha$. 
This result confirms that interactions modify the concrete form of the zero modes and the effective diagonal bases of the braiding, but not affecting the SPNA statistics since the generic symmetry protection mechanism is preserved.

\section{Conclusions and Discussions}\label{SecEnd}

In summary, we have shown that the PZMs with unitary symmetry obey the symmetry-protected non-Abelian (SPNA) statistics, manifesting a new paradigm of quantum statistics in strongly correlated systems, and established a systematic theory. The fractional topological phases hosting unitary symmetry-protected PZMs are strongly correlated and generically cannot be characterized through decoupled symmetry sectors, in sharp contrast to the free-fermion topological superconductors with symmetry-protected MZMs. Nevertheless, we have shown that the PZMs protected by unitary symmetry can always be classified into nontrivial symmetry sectors, with each sector preserving anyon parity individually in the dynamical evolutions. This profound symmetry protection mechanism determines the generic properties of the effective braiding Hamiltonian and, together with the anyon spin-statistics theorem, further leads to the SPNA statistics of PZMs. Finally, we have proposed a concrete physical model based on interacting nanowires to realized the PZMs with mirror symmetry, which satisfies the generic theory of the SPNA statistics.

This study advances the SPNA statistics from free-fermion topological states to a new realm of strongly correlated topological phases whose classifications are much broader~\cite{Wen2017Zoo}. Our prediction broadly expands the basic notion of non-Abelian statistics in combining with symmetry protection and strong correlations. A lot of future important issues deserve further in-depth study. For instance, while in the present study we have focused on the PZMs protected by Abelian unitary symmetries, extending the results to the scenarios with non-Abelian symmetries, which can protect coexisting PZMs more than two, may generate fundamentally new type of SPNA statistics due to the nonlinear features in the fractional quasiparticles, possibly achieving the universal quantum computation. On the other hand, extending SPNA statistics to higher-dimensional non-Abelian topological states~\cite{CChan2017,ZCGu2021} may yield the symmetry-protected non-Abelian anyons in beyond one and two dimensions.

The roles of symmetries in quantum statistics may revolutionize the basic understanding of quantum many-body physics. The present SPNA statistics for PZMs may be naturally applied to introduce new fractional Abelian statistics by considering symmetries. Generally, when there are multiple locally coexisting fractionalized particles, with or without the symmetry protection, those fractional particles may obey fundamentally different quantum statistics. A novel example may be given for semions. Without symmetry protection, it is well-known that semion pairs obey boson statistics, which was a key motivation to propose the anyon superconductors~\cite{WilczekBook,WittenAnyonSC,FisherAnyonSC,LaughlinAnyonSC}. However, if there is symmetry protection, the semion pairs formed from two symmetry sectors may satisfy the fermion rather than boson statistics. These are highly novel topics for future research.


\section*{ACKNOWLEDGMENTS}
This work was supported by National Key Research and Development Program of China (2021YFA1400900), the National Natural Science Foundation of China (Grants No. 11825401 and No. 12261160368), the Innovation Program for Quantum Science and Technology (Grant No. 2021ZD0302000), and the Strategic Priority Research Program of the Chinese Academy of Science (Grant No. XDB28000000).

\noindent


\indent

\appendix

\section{Unitary symmetry protection condition}\label{appenA}

We show that the generic unitary symmetry protection condition for a pair of PZMs can be expressed by the action of the unitary symmetry at the local parity operator. By definition, the most general symmetry protection of these PZMs is defined by $[H_{\text{c}},S]\neq0$, for $H_{\text{c}}$ in any form, where $S$ is a symmetry of the system. To facilitate the derivaition, we work in the bases of eigenstates of local parity operator $\tilde{Q}_{i}$, where $\tilde{Q}_{i}=U$, $\tilde{\alpha}_{i}=\omega^{(N-1)/2}UV$, $\tilde{\beta}_{i}=V$, and $U|q\rangle=\omega^{q}|q\rangle$, $V|q\rangle=|q-1\rangle$ are the $Z$-gate and $X$-gate (shift matrix) in the $N$-qudit formalism. The local coupling is diagonal in this bases $H_{{\rm c}}={\rm diag}(h_{1},h_{2},\cdots,h_{N})$, where $h_{i}$ is real for hermitianity. We further assume that $S|q\rangle$ is still a local parity eigenstate, $S|q\rangle=e^{i\theta(q')}|q'\rangle$, which can be regarded as a kind of parity superselection rule. The general symmetry protection condition is equivalent to $[H_{{\rm c}},S]=0\Rightarrow H_{{\rm c}}={\rm const.}$, and we have
\begin{align}
(SH_{{\rm c}})_{ij}= & \sum S_{ik}H_{kj}=S_{ij}h_{j},\\
(H_{{\rm c}}S')_{ij}= & \sum H_{ik}S_{kj}=h_{i}S_{ij},
\end{align}
then an $n$-cycle in $S$ yields $n$ identical $h_{i}$ in $H_{{\rm c}}$ and vice versa. Thus we conclude that the unitary symmetry $S$ must be an $N$-cycle. In this case, there exists a local unitary transformation such that
\begin{equation}
GSG^{-1}=e^{i\vartheta/N}V^{p},\label{Xk}
\end{equation}
where $p$ and $N$ are coprime and $G$ commutes with the Hamiltonian of the whole system. The action of $G$ rearranges the eigenstates such that $S|q'\rangle=e^{i\theta(q)}|q'-p\rangle$ where $\sum\theta(q)=\vartheta$ and gauges the phase $\theta(q)$ by sending $|q'\rangle$ to  $|q''\rangle=e^{-i[\theta(1)+\theta(2)+\cdots+\theta(q')-q'\vartheta/N]}|q'\rangle$, then $S|q''\rangle=e^{i\vartheta/N}|q''-p\rangle$. Thus we have proved the Eq.~(\ref{Xk}), which can be rewritten in the operator form
\begin{equation}
SQ_{i}S^{-1}=\omega^{p}Q_{i},\label{UnitarySA}
\end{equation}
where $Q_i=G^{-1} \tilde{Q}_i G$ is a new local parity operator. An example can be given in $\mathbb{Z}_4$ case where $S \tilde{Q}_i S^{-1}=[-\tilde{Q}_i-(1+i)\tilde{Q}_i^2+i\tilde{Q}_i^3]/2$. In the eigen-bases of $\tilde{Q}_{i}$, $S|1\rangle=|3\rangle$, $S|3\rangle=|2\rangle$, $S|2\rangle=|4\rangle$, $S|4\rangle=|1\rangle$. The local unitary transformation that swaps $|2\rangle$ and $|3\rangle$ is then $G=1/2+[(1-i)\tilde{Q}_{i}+(1+i)\tilde{Q}_{i}^{3}]/4+S(-1+i\tilde{Q}_{i}-\tilde{Q}_{i}^2-i\tilde{Q}_{i}^3)/4-S^{2}/2+S^{3}(-1-\tilde{Q}_{i}+\tilde{Q}_i^{2}-\tilde{Q}_{i}^{3})/4$ and the new local parity operator has the property $SQ_iS^{-1}=e^{3\pi i/2}Q_i$. The Eq.~(\ref{UnitarySA}) suggests that the symmetry can advance the local parity by $p$-increments where $p$ and $N$ are coprime. For $N=2$, this means that the symmetry operation switches odd/even fermion parity to even/odd one.

\section{Symmetry operation on each of the PZMs}\label{appenB}

We show in this Appendix how the symmetry acts on the PZMs given that $SQ_{i}S^{-1}=\omega^{p}Q_{i}$. Generically under the unitary symmetry $S$, we have
\begin{equation}
S\alpha'_{i}S^{-1}=e^{i\theta}\alpha'_{i}\sum_{n=1}^{N-1}\lambda_{n}Q_{i}^{n},\label{actionalpha}
\end{equation}
where $\alpha'_i=G^{-1} \tilde{\alpha}_i G$, and it follows that
$S\beta'_{i}S^{-1}=e^{i\theta}\omega^{-p}\beta'_{i}\sum_{n=1}^{N-1}\lambda_{n}Q_{i}^{n}$ with $\beta'_i=G^{-1} \tilde{\beta}_i G$. The parameters $e^{i\theta}$ and $\lambda_{n}$ are constraint by the symmetry $S$. That is, since $S$ is a unitary operation, the algebraic relations are kept under this operation, namely
\begin{align}
1&=\left(S\alpha'_{i}S^{-1}\right)^{N}=\left(S\beta'_{i}S^{-1}\right)^{N},\label{nfold}\\
1&=S{\alpha'}_{i}^{\dagger}S^{-1}S\alpha'_{i}S^{-1}=S{\beta'}_{i}^{\dagger}S^{-1}S\beta'_{i}S^{-1}.\label{unitarity}
\end{align}
From Eq. (\ref{actionalpha}) and Eq. (\ref{unitarity}), we find that $\sum_{n=1}^{N-1}\lambda_{n}Q_{i}^{n}$ is unitary and can be replaced by $e^{i\sum_{n=1}^{N-1}\mu_{n}Q_{i}^{n}}$, where $\mu_{n}^{*}=\mu_{N-n}$. Together with Eq. (\ref{nfold}), we have $\theta=2\pi p_1/N$, where $p_1$ is an integer. Collecting all the facts, we find the action of $S$ on the PZMs can be written as
\begin{align}
S\alpha'_{i}S^{-1}= & \omega^{p_{1}}\alpha'_{i}e^{i\sum_{n=1}^{N-1}\mu_{n}Q_{i}^{n}},\\
S\beta'_{i}S^{-1}= & \omega^{p_{2}}\beta'_{i}e^{i\sum_{n=1}^{N-1}\mu_{n}Q_{i}^{n}},
\end{align}
where $p_{1}-p_{2}=p$, $\mu_{n}^{*}=\mu_{N-n}$. In the case of MZM where $N=2$, these equations are simply $S\alpha'_{i}S^{-1} = \pm\alpha'_i e^{i\mu_1 Q_i} =\pm(\cos\mu_{1}\alpha'_{i}+\sin\mu_{1}\beta'_{i})$ and $S\beta'_{i}S^{-1} =\mp\beta'_i e^{i\mu_{1} Q_i}= \mp(-\sin\mu_{1}\alpha'_{i}+\cos\mu_{1}\beta'_{i})$, {\it i.e.}, each MZM is transformed by the symmetry operator into a linear combination of the Majorana doublet.

\section{Mixing of two sectors in the new PZMs under perturbation}\label{appenC}

In this Appendix we show how the specific symmetric perturbation $H_p$ in Eq.~(\ref{latticeH}) mix the two sectors. Suppose that $f/J\ll1$, $V/J\ll1$ and $f/V\sim\mathcal{O}(1)$, the PZM can be constructed perturbatively as
\begin{equation}
\bar{\alpha}_{1}=\alpha_{1}+\sum_{i}m_{i}\hat{O}_{i}^{(1)}+\sum_{j}n_{j}\hat{O}_{j}^{(2)}+\cdots,
\end{equation}
where $m_{i}\sim\mathcal{O}(f/J)\sim\mathcal{O}(V/J)$ and $n_{j}\sim\mathcal{O}(f^{2}/J^{2})\sim\mathcal{O}(V^{2}/J^{2})\sim\mathcal{O}(fV/J^{2})$. To the zeroth order, we have $\tilde{\alpha}_{1}=\alpha_{1}$ and the following commutator
\begin{equation}
[H_{\text{latt}},\alpha_{1}]=(\omega^{*}-1)f\alpha_{2}+(1-\omega^{*})f^{*}\alpha_{2}^{\dagger}\alpha_{1}^{2}.
\end{equation}
To cancel out the term $(\omega^{*}-1)f\alpha_{2}$, we only have two choices $\alpha_{3}$ and $\alpha_{2}^{2}\alpha_{3}^{\dagger}$ in the first order perturbation, thus at least one of them is involved in $\sum_{i}m_{i}\hat{O}_{i}^{(1)}$. They give the following commutators
\begin{align}
[H_{\text{latt}},\alpha_{3}] & =  \hat{C}_1(\alpha)+\hat{C}_2(\alpha,\beta),\\
[H_{\text{latt}},\alpha_{2}^{2}\alpha_{3}^{\dagger}] & =  \hat{D}_1(\alpha)+\hat{D}_2(\alpha)+\hat{D}_3(\alpha,\beta),
\end{align}
where $\hat{C}_1(\alpha)=(1-\omega)J\alpha_{2}^{\dagger}\alpha_{3}^{2}+(\omega-1)J^{*}\alpha_{2}$, $\hat{C}_2(\alpha,\beta)=(1-\omega)V\alpha_{2}^{\dagger}\alpha_{3}^{2}\beta_{2}^{\dagger}\beta_{3}+(\omega-1)V^{*}\alpha_{2}\beta_{3}^{\dagger}\beta_{2}$ and $\hat{D}_1(\alpha)=(\omega^{-2}-\omega^{*})J\alpha_{2}+(\omega^{-3}-\omega^{-4})J^{*}\alpha_{3}^{-2}\alpha_{2}^{3}$, $\hat{D}_2(\alpha)=(1-\omega^{2})f\alpha_{1}^{\dagger}\alpha_{2}^{3}\alpha_{3}^{\dagger}+(1-\omega^{-2})f^{*}\alpha_{3}^{\dagger}\alpha_{2}\alpha_{1}$, $\hat{D}_3(\alpha,\beta)=(\omega^{-2}-\omega^{*})V\alpha_{2}\beta_{2}^{\dagger}\beta_{3}+(\omega^{-3}-\omega^{-4})V^{*}\alpha_{3}^{-2}\alpha_{2}^{3}\beta_{3}^{\dagger}\beta_{2}$. For each commuator, there are terms with $\beta_{i}$, then we are left with two possibilities: (i) the terms with $\beta_{i}$ are canceled out in the first order perturbation $[H_{p},\hat{O}_{i}^{(1)}]$, then one can check that certain terms with $\beta_{i}$ like $\alpha_{1}\beta_{3}^{\dagger}\beta_{2}$ or $\alpha_{1}^{\dagger}\alpha_{2}^{2}\beta_{3}^{\dagger}\beta_{2}$ or $\alpha_{2}^{2}\alpha_{3}^{\dagger}\beta_{2}^{2}\beta_{3}^{-2}$ are involved in $\hat{O}_{i}^{(1)}$; (ii) the terms with $\beta_{i}$ are canceled out in the second order perturbation $[H_{\text{latt}}-H_{p},\hat{O}_{j}^{(2)}]$, then certain terms with $\beta_{i}$ like $\alpha_{2}^{2}\alpha_{3}^{\dagger}\beta_{2}\beta_{3}^{\dagger}$ or $\alpha_{3}\beta_{3}^{\dagger}\beta_{2}$ are involved in $\hat{O}_{j}^{(2)}$. In conclusion, $\tilde{\alpha}_{1}$ cannot be constructed solely by terms with $\alpha_{i}$ , and its dynamics is governed by the full Hamiltonian rather than just $H_{\alpha}$. Note that in the above iteration procedure, $\beta_{j}^{\dagger}$ and $\beta_{i}$ always co-occur in $\hat{O}_{i}^{(n)}$, and $S\beta_{j}^{\dagger}\beta_{i}S^{-1}=\omega^{*}\omega\beta_{j}^{\dagger}\beta_{i}=\beta_{j}^{\dagger}\beta_{i}$, we find $\bar{\alpha}_{1}$ still eigenmode of the symmetry $S\bar{\alpha}_{1}S^{-1}=\bar{\alpha}_{1}$.

\noindent


\begin{thebibliography}{55}%
\makeatletter
\providecommand \@ifxundefined [1]{%
 \@ifx{#1\undefined}
}%
\providecommand \@ifnum [1]{%
 \ifnum #1\expandafter \@firstoftwo
 \else \expandafter \@secondoftwo
 \fi
}%
\providecommand \@ifx [1]{%
 \ifx #1\expandafter \@firstoftwo
 \else \expandafter \@secondoftwo
 \fi
}%
\providecommand \natexlab [1]{#1}%
\providecommand \enquote  [1]{``#1''}%
\providecommand \bibnamefont  [1]{#1}%
\providecommand \bibfnamefont [1]{#1}%
\providecommand \citenamefont [1]{#1}%
\providecommand \href@noop [0]{\@secondoftwo}%
\providecommand \href [0]{\begingroup \@sanitize@url \@href}%
\providecommand \@href[1]{\@@startlink{#1}\@@href}%
\providecommand \@@href[1]{\endgroup#1\@@endlink}%
\providecommand \@sanitize@url [0]{\catcode `\\12\catcode `\$12\catcode
  `\&12\catcode `\#12\catcode `\^12\catcode `\_12\catcode `\%12\relax}%
\providecommand \@@startlink[1]{}%
\providecommand \@@endlink[0]{}%
\providecommand \url  [0]{\begingroup\@sanitize@url \@url }%
\providecommand \@url [1]{\endgroup\@href {#1}{\urlprefix }}%
\providecommand \urlprefix  [0]{URL }%
\providecommand \Eprint [0]{\href }%
\providecommand \doibase [0]{http://dx.doi.org/}%
\providecommand \selectlanguage [0]{\@gobble}%
\providecommand \bibinfo  [0]{\@secondoftwo}%
\providecommand \bibfield  [0]{\@secondoftwo}%
\providecommand \translation [1]{[#1]}%
\providecommand \BibitemOpen [0]{}%
\providecommand \bibitemStop [0]{}%
\providecommand \bibitemNoStop [0]{.\EOS\space}%
\providecommand \EOS [0]{\spacefactor3000\relax}%
\providecommand \BibitemShut  [1]{\csname bibitem#1\endcsname}%
\let\auto@bib@innerbib\@empty



\bibitem{Shankarbook} R. Shankar, \emph{Principles of Quantum Mechanics}, (Springer, New York, 1994).

\bibitem{Anyon1977} J. M. Leinaas and J. Myrheim, \emph{On the theory of identical particles}, Nuovo Cimento Soc. Ital. Fis., B {\bf 37}, 1 (1977).

\bibitem{Wilczek1982} F. Wilczek, \emph{Magnetic Flux, Angular Momentum, and Statistics}, Phys. Rev. Lett. {\bf 48}, 1144 (1982).

\bibitem{WilczekBook} F. Wilczek, \emph{Fractional Statistics and Anyon Superconductivity} (World Scientific, Singapore, 1990).


\bibitem{Nayak1996} C. Nayak and F. Wilczek, \emph{$2n$-quasihole states realize $2^{n-1}$-dimensional spinor braiding statistics in paired quantum Hall states}, Nucl. Phys. B {\bf 479}, 529 (1996).

\bibitem{Ivanov2001} D. A. Ivanov, \emph{Non-Abelian Statistics of Half-Quantum Vortices in $p$-Wave Superconductors}, Phys. Rev. Lett. {\bf 86}, 268 (2001).

\bibitem{DasSarma2005} S. Das Sarma, M. Freedman, and C. Nayak, \emph{Topologically protected qubits from a possible non-Abelian fractional quantum Hall state}, Phys. Rev. Lett. {\bf 94}, 166802 (2005).

\bibitem{Nayak2008} C. Nayak, S. H. Simon, A. Stern, M. Freedman, and S. Das Sarma, \emph{Non-Abelian anyons and topological quantum computation}, Rev. Mod. Phys. {\bf 80}, 1083 (2008).

\bibitem{Alicea2011} J. Alicea, Y. Oreg, G. Refael, F. Von Oppen, and M. P. Fisher, \emph{Non-Abelian statistics and topological quantum information processing in 1D wire networks}, Nat. Phys. {\bf 7}, 412 (2011).


\bibitem{Berry1997} M. V. Berry and J. M. Robbins, \emph{Indistinguishability for quantum particles: spin, statistics and the geometric phase}, Proc. R. Soc. London A {\bf 453}, 1771 (1997).

\bibitem{GeometricNRP2019} E. Cohen, H. Larocque, F. Bouchard, F. Nejadsattari, Y. Gefen, and E. Karimi, \emph{Geometric phase from Aharonov-Bohm to Pancharatnam-Berry and beyond}, Nat. Rev. Phys. {\bf 1}, 437 (2019).


\bibitem{SternReview2010} A. Stern, \emph{Non-Abelian states of matter}, Nature (London) {\bf 464}, 187 (2010).


\bibitem{Kitaev2003} A. Kitaev, \emph{Fault-Tolerant Quantum Computation by Anyons}, Ann. Phys. (Amsterdam) {\bf 303}, 2 (2003).

\bibitem{PachosBook} J. K. Pachos, \emph{Introduction to Topological Quantum Computation} (Cambridge University Press, Cambridge, England, 2012).

\bibitem{Pachos2017} V. Lahtinen and J. K. Pachos, \emph{A Short Introduction to Topological Quantum Computation}, SciPost Phys. {\bf 3}, 021 (2017).


\bibitem{Mourik2012} V. Mourik, K. Zuo, S. M. Frolov, S. Plissard, E. P. Bakkers, and L. P. Kouwenhoven, \emph{Signatures of Majorana fermions in hybrid superconductor-semiconductor nanowire devices}, Science {\bf 336}, 1003 (2012).

\bibitem{MTDeng2012} M. Deng, C. Yu, G. Huang, M. Larsson, P. Caroff, and H. Xu, \emph{Anomalous zero-bias conductance peak in a Nb-InSb nanowire-Nb hybrid device}, Nano Letters {\bf 12}, 6414 (2012).

\bibitem{Rokhinson2012} L. P. Rokhinson, X. Liu, and J. K. Furdyna, \emph{The fractional ac Josephson effect in a semiconductor-superconductor nanowire as a signature of Majorana particles}, Nat. Phys. {\bf 8}, 795 (2012).

\bibitem{Shtrikman2012} A. Das, Y. Ronen, Y. Most, Y. Oreg, M. Heiblum, and H. Shtrikman, \emph{Zero-bias peaks and splitting in an Al-InAs nanowire topological superconductor as a signature of Majorana fermions}, Nat. Phys. {\bf 8}, 887 (2012).

\bibitem{JFJia2012} M.-X. Wang, C. Liu, J.-P. Xu, F. Yang, L. Miao, M.-Y. Yao, C. L. Gao, C. Shen, X. Ma, X. Chen, Z.-A. Xu, Y. Liu, S.-C. Zhang, D. Qian, J.-F. Jia, and Q.-K. Xue, \emph{The Coexistence of Superconductivity and Topological Order in the $Bi_2Se_3$ Thin Films}, Science {\bf 336}, 52 (2012).

\bibitem{Marcus2013} H. O. H. Churchill, V. Fatemi, K. Grove-Rasmussen, M. T. Deng, P. Caroff, H. Q. Xu, and C. M. Marcus, \emph{Superconductor-nanowire devices from tunneling to the multichannel regime: Zero-bias oscillations and magnetoconductance crossover}, Phys. Rev. B {\bf 87}, 241401 (2013).

\bibitem{JFJia2014} J.-P. Xu, C. Liu, M.-X. Wang, J. Ge, Z.-L. Liu, X. Yang, Y. Chen, Y. Liu, Z.-A. Xu, C.-L. Gao, D. Qian, F.-C. Zhang, and J.-F. Jia, \emph{Artificial topological superconductor by the proximity effect}, Phys. Rev. Lett. {\bf 112}, 217001 (2014).

\bibitem{Yazdani2014} S. Nadj-Perge, I. K. Drozdov, J. Li, H. Chen, S. Jeon, J. Seo, A. H. MacDonald, B. A. Bernevig, and A. Yazdani, \emph{Observation of Majorana fermions in ferromagnetic atomic chains on a superconductor}, Science {\bf 346}, 602 (2014).

\bibitem{Marcus2015} W. Chang, S. Albrecht, T. Jespersen, F. Kuemmeth, P. Krogstrup, J. Nyg{\aa}rd, and C. M. Marcus, \emph{Hard gap in epitaxial semiconductor-superconductor nanowires}, Nature Nanotechnology {\bf 10}, 232 (2015).

\bibitem{Marcus2016} S. M. Albrecht, A. P. Higginbotham, M. Madsen, F. Kuemmeth, T. S. Jespersen, J. Nyg{\aa}rd, P. Krogstrup, and C. Marcus, \emph{Exponential protection of zero modes in Majorana islands}, Nature {\bf 531}, 206 (2016).

\bibitem{Molenkamp2016} J. Wiedenmann, E. Bocquillon, R. S. Deacon, S. Hartinger, O. Herrmann, T. M. Klapwijk, L. Maier, C. Ames, C. Br{\"u}ne, C. Gould, A. Oiwa, K. Ishibashi, S. Tarucha, H. Buhmann, and L. W. Molenkamp, \emph{ 4$\pi$-periodic Josephson supercurrent in HgTe-based topological Josephson junctions}, Nature Communications {\bf 7}, 1 (2016).

\bibitem{Molenkamp2017} E. Bocquillon, R. S. Deacon, J. Wiedenmann, P. Leubner, T. M. Klapwijk, C. Br{\"u}ne, K. Ishibashi, H. Buhmann, and L. W. Molenkamp, \emph{Gapless Andreev bound states in the quantum spin Hall insulator HgTe}, Nature Nanotechnology {\bf 12}, 137 (2017).

\bibitem{HDing2018a} P. Zhang, K. Yaji, T. Hashimoto, Y. Ota, T. Kondo, K. Okazaki, Z. Wang, J. Wen, G. D. Gu, H. Ding, and S. Shin, \emph{Observation of topological superconductivity on the surface of an iron-based superconductor}, Science {\bf 360}, 182 (2018).

\bibitem{HDing2018b} D. Wang, L. Kong, P. Fan, H. Chen, S. Zhu, W. Liu, L. Cao, Y. Sun, S. Du, J. Schneeloch, R. Zhong, G. Gu, L. Fu, H. Ding, and H.-J. Gao, \emph{Evidence for Majorana bound states in an iron-based superconductor}, Science {\bf 362}, 333 (2018).

\bibitem{Marcus2019} A. Fornieri, A. M. Whiticar, F. Setiawan, E. Portol{\'e}s, A. C. Drachmann, A. Keselman, S. Gronin, C. Thomas, T. Wang, R. Kallaher, G. C. Gardner, E. Berg, M. J. Manfra, A. Stern, C. M. Marcus, and F. Nichele, \emph{Evidence of topological superconductivity in planar Josephson junctions}, Nature {\bf 569}, 89 (2019).

\bibitem{Molenkamp2019} H. Ren, F. Pientka, S. Hart, A. T. Pierce, M. Kosowsky, L. Lunczer, R. Schlereth, B. Scharf, E. M. Hankiewicz, L. W. Molenkamp, B. I. Halperin, and A. Yacoby, \emph{Topological superconductivity in a phase-controlled Josephson junction}, Nature {\bf 569}, 93 (2019).

\bibitem{Yazdani2019} B. J{\aa}ck, Y. Xie, J. Li, S. Jeon, B. A. Bernevig, and A. Yazdani, \emph{Observation of a Majorana zero mode in a topologically protected edge channel}, Science {\bf 364}, 1255 (2019).

\bibitem{Zhang2022} Z. Wang, H. Song, D. Pan, Z. Zhang, W. Miao, R. Li, Z. Cao, G. Zhang, L. Liu, L. Wen, R. Zhuo, D. E. Liu, K. He, R. Shang, J. Zhao, and H. Zhang, \emph{Plateau Regions for Zero-Bias Peaks within 5\% of the Quantized Conductance Value $2e^2/h$}, Phys. Rev. Lett. {\bf 129}, 167702 (2022).

\bibitem{Microsoft2023} M. Aghaee, A. Akkala, Z. Alam, R. Ali, A. A. Ramirez, M. Andrzejczuk, A. E. Antipov, M. Astafev, B. Bauer, J. Becker
\emph{et al.}, \emph{InAs-Al hybrid devices passing the topological gap protocol}, Phys. Rev. B {\bf 107}, 245423 (2023).


\bibitem{Kitaev2001} A. Y. Kitaev, \emph{Unpaired Majorana Fermions in Quantum Wires}, Phys. Usp. {\bf 44}, 131 (2001).

\bibitem{AliceaRPP} J. Alicea, \emph{New directions in the pursuit of Majorana fermions in solid state systems}, Rep. Prog. Phys. {\bf 75}, 076501 (2012).

\bibitem{Flensberg2012} M. Leijnse and K. Flensberg, \emph{Introduction to topological superconductivity and Majorana fermions}, Semicond. Sci. Technol. {\bf 27}, 124003 (2012).

\bibitem{Beenakker2013} C. W. J. Beenakker, \emph{Search for Majorana fermions in superconductors}, Annu. Rev. Condens. Matter Phys. 4, {\bf 113} (2013).

\bibitem{Franz2015} S. R. Elliott and M. Franz, \emph{Colloquium: Majorana fermions in nuclear, particle, and solid-state physics}, Rev. Mod. Phys. {\bf 87}, 137 (2015).

\bibitem{Sato2017} M. Sato and Y. Ando, \emph{Topological superconductors: A review}, Rep. Prog. Phys. {\bf 80}, 076501 (2017).

\bibitem{YPHe2020} Y.-P. He, J.-S. Hong, and X.-J. Liu, \emph{Non-abelian statistics of Majorana modes and the applications to topological quantum computation}. Acta Phys. Sin. {\bf 69}, 110302 (2020).


\bibitem{Gottesman1998} D. Gottesman, \emph{The Heisenberg representation of quantum computers}, arXiv:quant-ph/9807006 (1998)


\bibitem{FradkinKadanoff} E. Fradkin and L. P. Kadanoff, \emph{Disorder variables and para-fermions in two-dimensional statistical mechanics}, Nucl. Phys. B {\bf 170}, 1 (1980).

\bibitem{Fendley2012} P. Fendley, \emph{Parafermionic edge zero modes in $\mathbb{Z}_n$-invariant spin chains}, J. Stat. Mech. (2012) P11020.

\bibitem{Motruk2013topological} J. Motruk, E. Berg, A. M. Turner, and F. Pollmann, \emph{Topological Phases in Gapped Edges of Fractionalized
Systems}, Phys. Rev. B {\bf 88}, 085115 (2013).

\bibitem{Bondesan2013} R. Bondesan and T. Quella, \emph{Topological and symmetry broken phases of $\mathbb{Z}_N$ parafermions in one dimension}, J. Stat. Mech. (2013) P10024.

\bibitem{Meidan2017} D. Meidan, E. Berg, and A. Stern, \emph{Classification of Topological Phases of Parafermionic Chains with Symmetries}, Phys. Rev. B {\bf 95}, 205104 (2017).


\bibitem{GMZhang2018a} W.-T. Xu and G.-M. Zhang, \emph{Classifying parafermionic gapped phases using matrix product states}, Phys. Rev. B {\bf 97}, 035160 (2018).


\bibitem{SternPRX2012} N. H. Lindner, E. Berg, G. Refael, and A. Stern, \emph{Fractionalizing Majorana Fermions: Non-Abelian Statistics on the Edges of Abelian Quantum Hall States}, Phys. Rev. X {\bf 2}, 041002 (2012).

\bibitem{MCheng2012} M. Cheng, \emph{Superconducting proximity effect on the edge of fractional topological insulators}, Phys. Rev. B {\bf 86}, 195126 (2012).

\bibitem{AliceaNC2013} D. J. Clarke, and J. Alicea, \emph{Exotic non-Abelian anyons from conventional fractional quantum Hall states}, Nat. Commun. {\bf 4}, 1348 (2013).

\bibitem{Barkeshli2013} M. Barkeshli, C.-M. Jian, and X.-L. Qi, \emph{Twist Defects and Projective Non-Abelian Braiding Statistics}, Phys. Rev. B {\bf 87}, 045130 (2013).

\bibitem{Barkeshli2014} M. Barkeshli and X.-L. Qi, \emph{Synthetic Topological Qubits in Conventional Bilayer Quantum Hall Systems}, Phys. Review X {\bf 4}, 041035 (2014).

\bibitem{Stern2014} Y. Oreg, E. Sela, and A. Stern, \emph{Fractional helical liquids in quantum wires}, Phys. Rev. B {\bf 89}, 115402 (2014).

\bibitem{Loss2014a} J. Klinovaja and D. Loss, \emph{Parafermions in an Interacting Nanowire Bundle}, Phys. Rev. Lett. {\bf 112}, 246403 (2014).

\bibitem{Loss2014b} J. Klinovaja and D. Loss, \emph{Time-Reversal Invariant Parafermions in Interacting Rashba Nanowires}, Phys. Rev. B {\bf 90}, 045118 (2014).

\bibitem{Loss2014c} J. Klinovaja, A. Yacoby, and D. Loss, \emph{Kramers Pairs of Majorana Fermions and Parafermions in Fractional Topological Insulators}, Phys. Rev. B {\bf 90}, 155447 (2014).

\bibitem{Oreg2017general} H. Ebisu, E. Sagi, Y. Tanaka, and Y. Oreg, \emph{Generalized parafermions and nonlocal Josephson effect in multilayer systems}, Phys. Rev. B {\bf 95}, 075111 (2017).

\bibitem{Oreg2017chiral} E. Sagi, A. Haim, E. Berg, F. von Oppen, and Y. Oreg, \emph{Fractional chiral superconductors}, Phys. Rev. B {\bf 96}, 235144 (2017).

\bibitem{Loss2019} K. Laubscher, D. Loss, and J. Klinovaja, \emph{Fractional topological superconductivity and parafermion corner states}, Phys. Rev. Research {\bf 1}, 032017(R) (2019).

\bibitem{Santos2020} L. H. Santos, \emph{Parafermions in hierarchical fractional quantum Hall states}, Phys. Rev. Research {\bf 2}, 013232 (2020).

\bibitem{Loss2020bilayer} K. Laubscher, D. Loss, and J. Klinovaja, \emph{Majorana and parafermion corner states from two coupled sheets of bilayer graphene}, Phys. Rev. Research {\bf 2}, 013330 (2020).

\bibitem{Oreg2020predicted} N. Schiller, E. Cornfeld, E. Berg, and Y. Oreg, \emph{Predicted Signatures of Topological Superconductivity and Parafermion Zero Modes in Fractional Quantum Hall Edges}, Phys. Rev. Research {\bf 2}, 023296 (2020).

\bibitem{Silva2022} R. L. R. C. Teixeira and L. G. G. V. Dias da Silva, \emph{Edge $\mathbb{Z}_3$ parafermions in fermionic lattices}, Phys. Rev. B {\bf 105}, 195121 (2022).

\bibitem{Gefen2022} U. Khanna, M. Goldstein, and Y. Gefen, \emph{Parafermions in a multilegged geometry: Towards a scalable  parafermionic network}, Phys. Rev. B {\bf 105}, L161101 (2022).

\bibitem{FockPara} E. Cobanera and G. Ortiz, \emph{Fock Parafermions and Self-Dual Representations of the Braid Group}, Phys. Rev. A {\bf 89}, 012328 (2014).

\bibitem{Loss2016} A. Hutter and D. Loss, \emph{Quantum Computing with Parafermions}, Phys. Rev. B {\bf 93}, 125105 (2016).


\bibitem{SCZhang2009} X.-L. Qi, T. L. Hughes, S. Raghu, and S.-C. Zhang, \emph{Time-reversal-invariant topological superconductors and superfluids in two and three dimensions}, Phys. Rev. Lett. {\bf 102}, 187001 (2009).

\bibitem{TeoKane2010} J. C. Y. Teo and C. L. Kane, \emph{Topological defects and gapless modes in insulators and superconductors}, Phys. Rev. B {\bf 82}, 115120 (2010).

\bibitem{Timm2010} A. P. Schnyder, P. M. R. Brydon, D. Manske, and C. Timm, \emph{Andreev spectroscopy and surface density of states for a three-dimensional time-reversal-invariant topological superconductor}, Phys. Rev. B {\bf 82}, 184508 (2010).

\bibitem{Beenakker2011} C. W. J. Beenakker, J. P. Dahlhaus, M. Wimmer, and A. R. Akhmerov, \emph{Random-matrix theory of Andreev reflection from a topological superconductor}, Phys. Rev. B {\bf 83}, 085413 (2011).

\bibitem{KTLaw2012} C. L. M. Wong and K. T. Law, \emph{Majorana Kramers doublets in $d_{x^2-y^2}$-wave superconductors with Rashba spin-orbit coupling}, Phys. Rev. B {\bf 86}, 184516 (2012).

\bibitem{Nagaosa2012} S. Nakosai, Y. Tanaka, and N. Nagaosa, \emph{Topological superconductivity in bilayer Rashba system}, Phys. Rev. Lett. {\bf 108}, 147003 (2012).

\bibitem{KaneMele2013} F. Zhang, C. L. Kane, and E. J. Mele, \emph{Time-reversal-invariant topological superconductivity and Majorana Kramers pairs}, Phys. Rev. Lett. {\bf 111}, 056402 (2013).

\bibitem{Berg2013} A. Keselman, L. Fu, A. Stern, and E. Berg, \emph{Inducing time-reversal-invariant topological superconductivity and fermion parity pumping in quantum wires}, Phys. Rev. Lett. {\bf 111}, 116402 (2013).

\bibitem{DLoss-a} J. Klinovaja, A. Yacoby, and D. Loss, \emph{Kramers pairs of Majorana fermions and parafermions in fractional topological insulators}, Phys. Rev. B {\bf 90}, 155447 (2014).

\bibitem{DLoss-b} J. Klinovaja and D. Loss, \emph{Time-reversal invariant parafermions in interacting Rashba nanowires}, Phys. Rev. B {\bf 90}, 045118 (2014).

\bibitem{Nagaosa2014} R. Wakatsuki, M. Ezawa and N. Nagaosa, \emph{Majorana fermions and multiple topological phase transition in Kitaev ladder topological superconductors}, Phys. Rev. B {\bf 89}, 174514 (2014).

\bibitem{magneticTSC} C. Fang, M. J. Gilbert and B. A. Bernevig, \emph{New class of topological superconductors protected by magnetic group symmetries}, Phys. Rev. Lett. {\bf 112}, 106401 (2014).

\bibitem{Oreg2019} A. Haim and Y. Oreg, \emph{Time-reversal-invariant topological superconductivity in one and two dimensions}, Physics Reports {\bf 825}, 1 (2019).



\bibitem{mirrorTSC} F. Zhang, C. Kane and E. Mele, \emph{Topological mirror superconductivity}, Phys. Rev. Lett. {\bf 111}, 056403 (2013).

\bibitem{Ueno2013} Y. Ueno, A. Yamakage, Y. Tanaka and M. Sato, \emph{Symmetry-protected Majorana fermions in topological crystalline superconductors: theory and application to Sr2RuO4}, Phys. Rev. Lett. {\bf 111}, 087002 (2013).

\bibitem{C4TSC} X.-J. Liu, J. J. He and K. T. Law, \emph{Demonstrating lattice symmetry protection in topological crystalline superconductors}, Phys. Rev. B {\bf 90}, 235141 (2014).

\bibitem{Sato2014crystalline} K. Shiozaki and M. Sato, \emph{Topology of crystalline insulators and superconductors}, Phys. Rev. B {\bf 90}, 165114 (2014).

\bibitem{InversionTSC} Y.-T. Hsu, W. S. Cole, R.-X. Zhang, and J. D. Sau,  \emph{Inversion-protected higher-order topological superconductivity in monolayer WTe2}, Phys. Rev. Lett. {\bf 125}, 097001 (2020).


\bibitem{XJL2014PRX} X.-J. Liu, C. L. M. Wong, and K. T. Law, \emph{Non-Abelian Majorana Doublets in Time-Reversal-Invariant Topological Superconductors}, Phys. Rev. X {\bf 4}, 021018 (2014).

\bibitem{Gao2016} P. Gao, Y.-P. He, and X.-J. Liu, \emph{Symmetry-protected non-Abelian braiding of Majorana Kramers pairs}, Phys. Rev. B {\bf 94}, 224509 (2016); {\bf 95}, 019902(E) (2017).

\bibitem{Sato2014mirror} M. Sato, A. Yamakage, and T. Mizushima, \emph{Mirror Majorana zero modes in spinful superconductors/superfluids non-Abelian anyons in integer quantum vortices}, Physica E {\bf 55}, 20 (2014).

\bibitem{JSH2022} J.-S. Hong, T.-F. J. Poon, L. Zhang, and X.-J. Liu, \emph{Unitary symmetry-protected non-Abelian statistics of Majorana modes}, Phys. Rev. B {\bf 105}, 024503 (2022).

\bibitem{YWuReview2023} Y. Wu, J. Liu, and X. C. Xie, \emph{Recent progress on non-Abelian anyons: from Majorana zero modes to topological Dirac fermionic modes}, Sci. China Phys. Mech. Astron. {\bf 66}, 267004 (2023). https://doi.org/10.1007/s11433-022-2015-y

\bibitem{NittaReview2024} Y. Masaki, T. Mizushima, and M. Nitta, \emph{Non-abelian anyons and non-abelian vortices in topological superconductors}, in \emph{Encyclopedia of Condensed Matter Physics
(Second Edition)}, edited by T. Chakraborty (Academic
Press, Oxford, 2024) second edition, pp. 755-794. https://doi.org/10.1016/B978-0-323-90800-9.00225-0


\bibitem{Nitta2012dirac} S. Yasui, K. Itakura, and M. Nitta, \emph{Dirac returns: Non-abelian statistics of vortices with dirac fermions}, Nucl. Phys. B {\bf 859}, 261 (2012).

\bibitem{Loss2013fractional} J. Klinovaja and D. Loss, \emph{Fractional fermions with non-abelian statistics}, Phys. Rev. Lett. {\bf 110}, 126402 (2013).

\bibitem{YWu2020a} Y. Wu, H. Liu, J. Liu, H. Jiang, and X. C. Xie, \emph{Double-frequency Aharonov-Bohm effect and non-Abelian braiding properties of Jackiw-Rebbi zero-mode}, Natl. Sci. Rev. {\bf 7}, 572 (2020).

\bibitem{YWu2020b} Y. Wu, H. Jiang, J. Liu, H. Liu, and X. C. Xie, \emph{Non-Abelian braiding of Dirac fermionic modes using topological corner states in higher-order topological insulator}, Phys. Rev. Lett. {\bf 125}, 036801 (2020).

\bibitem{YWu2022} Y. Wu, H. Jiang, H. Chen, H. Liu, J. Liu, and X. C. Xie, \emph{Non-Abelian braiding in spin superconductors utilizing the Aharonov-Casher effect}, Phys. Rev. Lett. {\bf 128}, 106804 (2022).


\bibitem{localmixing1} K. W\"olms, A. Stern, and K. Flensberg, \emph{Local adiabatic mixing of kramers pairs of majorana bound states}, Phys. Rev. Lett. {\bf 113}, 246401 (2014).

\bibitem{localmixing2} K. W\"olms, A. Stern, and K. Flensberg, \emph{Braiding properties of Majorana Kramers pairs}, Phys. Rev. B {\bf 93}, 045417 (2016).

\bibitem{Knapp2020} C. Knapp, A. Chew, and J. Alicea, \emph{Fragility of the fractional josephson effect in time-reversal-invariant topological superconductors}, Phys. Rev. Lett. {\bf 125}, 207002 (2020).


\bibitem{JasonAlicea2020} C. Knapp, A. Chew, and J. Alicea, \emph{Fragility of the Fractional Josephson Effect in Time-Reversal-Invariant Topological Superconductors}, Phys. Rev. Lett. {\bf 125}, 207002 (2020).

\bibitem{Cooper2020} M. McGinley and N. R. Cooper, \emph{Fragility of time-reversal symmetry protected topological phases}, Nat. Phys. {\bf 16}, 1181 (2020).

\bibitem{Zhai2021} T.-S. Deng, L. Pan, Y. Chen, and H. Zhai, \emph{Stability of Time-Reversal Symmetry Protected Topological Phases}, Phys. Rev. Lett. {\bf 127}, 086801 (2021).

\bibitem{Cai2021} Z. Wang, Q. Li, W. Li, and Z. Cai, \emph{Symmetry-Protected Topological Edge Modes and Emergent Partial Time-Reversal Symmetry Breaking in Open Quantum Many-Body Systems}, Phys. Rev. Lett. {\bf 126}, 237201 (2021).


\bibitem{OregReview2019} A. Haim and Y. Oreg, \emph{Time-reversal-invariant topological superconductivity in one and two dimensions}, Phys. Rep. {\bf 825}, 1 (2019).

\bibitem{MCheng2021} M. F. Lapa, M. Cheng, and Y. Wang, \emph{Symmetry-protected gates of Majorana qubits in a high-$T_c$ higher-order topological superconductor platform}, SciPost Phys. {\bf 11}, 086 (2021).

\bibitem{TanakaMKP2022} Y. Tanaka, T. Sanno, T. Mizushima, and S. Fujimoto, \emph{Manipulation of Majorana-Kramers qubit and its tolerance in time-reversal invariant topological superconductor}, Phys. Rev. B {\bf 106}, 014522 (2022).

\bibitem{LFu2022} C. Schrade and L. Fu, \emph{Quantum Computing with Majorana Kramers Pairs}, Phys. Rev. Lett. {\bf 129}, 227002 (2022).


\bibitem{XGWen2012} X.-G. Wen, \emph{Symmetry-protected topological phases in noninteracting fermion systems}, Phy. Rev. B {\bf 85}, 085103 (2012).


\bibitem{Kitaev2006} A. Kitaev, \emph{Anyons in an exactly solved model and beyond}, Ann. Phys. (Amsterdam) {\bf 321}, 2 (2006).



\bibitem{Bonderson2008} P. Bonderson, M. Freedman, and C. Nayak, \emph{Measurement-only topological quantum computation}, Phys. Rev. Lett. {\bf 101}, 010501 (2008).


\bibitem{Giamarchi} T. Giamarchi, \emph{Quantum Physics in One Dimension}, (Oxford University Press, New York, 2004).



\bibitem{Wen2017Zoo} X.-G. Wen, \emph{Colloquium: Zoo of quantum-topological phases of matter}, Rev. Mod. Phys. {\bf 89}, 041004 (2017).




\bibitem{CChan2017} C. Chan and X. J. Liu, \emph{Non-Abelian Majorana Modes Protected by an Emergent Second Chern Number}, Phys. Rev. Lett. {\bf 118}, 207002 (2017).

\bibitem{ZCGu2021} J.-R. Zhou, Q.-R. Wang, C. Wang, and Z.-C. Gu, \emph{Non-Abelian three-loop braiding statistics for 3D fermionic topological
phases}, Nat. Commun. {\bf 12}, 3191 (2021).


\bibitem{WittenAnyonSC} Y.-H. Chen, F. Wilczek, E. Witten, and B. Halperin, \emph{On anyon superconductivity}, Int. J. Mod. Phys. B {\bf 3}, 1001 (1989).

\bibitem{FisherAnyonSC} D.-H. Lee and M. P. A. Fisher, \emph{Anyon Superconductivity and the Fractional Quantum Hall Effect}, Phys. Rev. Lett. {\bf 63}, 903 (1989).

\bibitem{LaughlinAnyonSC} R. B. Laughlin, \emph{Current Status of Semionic Pairing Theory of High-$T_c$ Superconductors}, Int. J. Mod. Phys. B {\bf 5}, 1507 (1991).

\end{thebibliography}
\end{document}